# Anomalous diffusion in the dynamics of complex processes

Serge F. Timashev,[1,2] Yuriy S. Polyakov,[3] Pavel I. Misurkin,[4] and Sergey G. Lakeev[1]

[1]*Karpov Institute of Physical Chemistry, Ul. Vorontsovo pole 10, Moscow 103064, Russia*

[2]*Institute of Laser and Information Technologies, Russian Academy of Sciences, Troitsk, Pionerskaya str. 2, Moscow Region, 142190, Russia*

[3]*USPolyResearch, Ashland, PA 17921, USA*

[4]*Semenov Institute of Physical Chemistry, Russian Academy of Sciences, Ul. Kosygina 4, Moscow 19991, Russia*

Anomalous diffusion, process in which the mean-squared displacement of system states is a non-linear function of time, is usually identified in real stochastic processes by comparing experimental and theoretical displacements at relatively small time intervals. This paper proposes an interpolation expression for the identification of anomalous diffusion in complex signals for the cases when the dynamics of the system under study reaches a steady state (large time intervals). This interpolation expression uses the chaotic difference moment (transient structural function) of the second order as an average characteristic of displacements. A general procedure for identifying anomalous diffusion and calculating its parameters in real stochastic signals, which includes the removal of the regular (low-frequency) components from the source signal and the fitting of the chaotic part of the experimental difference moment of the second order to the interpolation expression, is presented. The procedure was applied to the analysis of the dynamics of magnetoencephalograms, blinking fluorescence of quantum dots, and X-ray emission from accreting objects. For all three applications, the interpolation was able to adequately describe the chaotic part of the experimental difference moment, which implies that anomalous diffusion manifests itself in these natural signals. The results of this study make it possible to broaden the range of complex natural processes in which anomalous diffusion can be identified. The relation between the interpolation expression and a diffusion model, which is derived in the paper, allows one to simulate the chaotic processes in the open complex systems with anomalous diffusion. It is shown that the equations of anomalous diffusion with general integrodifferential boundary conditions should be used for modeling these processes.

## I. INTRODUCTION

Anomalous diffusion [1-10] is a "random walk" process, for which the average squared deviation of system states $V(t)$, varying during time $t$ over the whole set of possible states, $(-\infty < V < \infty)$, from the average value can be written as

$$\left\langle [\Delta V(t)]^2 \right\rangle_{pdf} = 2Dt_0 (t/t_0)^{2H_1}. \qquad (1)$$

Here, $D$ is the diffusion coefficient; $t_0$ is the characteristic time; $H_1$ is the Hurst constant; the averaging, denoted by the symbol $\langle...\rangle_{pdf}$, is effected by introducing the probability density function $W(V, t)$, which accounts for the probability of the system state being within the given interval of states at a specific moment $t$. It is assumed that the system was found in the vicinity of the $V = 0$ state (point) at the initial time $t = 0$. Fickian diffusion ($H_1 = 0.5$) corresponds to the random walks of system states characterized by some characteristic scale $\delta V$ of the values of elementary jumps, which are associated with the system transfers between adjacent states, and by the characteristic residence time $\delta \tau$ for every state. However, if these random walks of states stochastically alternate with the jumps having anomalous values higher than $\delta V$ at the same characteristic residence times $\delta \tau$ for the given state, the so-called superdiffusion (Lévy diffusion or Lévy flights), for which $H_1 > 0.5$, can occur. If the random walks stochastically alternate with the jumps having anomalously long times of residence in some states ("stability islands" [4]), which are much larger than $\delta\tau$, for the same values of jumps $\delta V$, the so-called subdiffusion, for which $H_1 < 0.5$, can occur.

Anomalous (from Fick's viewpoint) diffusion processes can be described using diffusion equations with constant diffusion coefficients in which the partial derivatives with respect to time and coordinate are replaced by fractional-order derivatives [1]. In this case, the subdiffusion process is described by introducing the fractional derivative of order $\alpha$ $(0<\alpha<1)$ instead of the partial derivative of the first order in time whereas the superdiffusion process is described by introducing the fractional derivative of order $\beta$ $(0<\beta<2)$ instead of the partial derivative of the second order with respect to coordinate [1-4]. The parameter $H_1$ is varied in the ranges $0 < H_1 < 0.5$ for subdiffusion and $0.5 < H_1 < 1$ for superdiffusion. It should be noted that if the process under study is more complicated, such as the one in which the diffusion coefficient depends on coordinate, the value of the Hurst constant can be higher than unity.

Presently, there is a substantial amount of data relating various random fluctuations to Eq. (1). Probably the most well-known example is the diffusion of particles in a turbulent flow, i.e., Richardson diffusion [11], which is described by Eq. (1) with $H_1$ = 1.5. There are examples of the occurrence of anomalous diffusion in the charge transfer in semiconductors, dynamics of biological and polymer systems, mass transfer in porous glass, and quantum optics [2, 3, 7-10].

This study shows that anomalous diffusion takes place in many other real processes. In particular, many natural signals represented as time series, variables $V(t)$ changing with time $t$, contain chaotic "random walk" components the dynamics of which can be described in terms of anomalous diffusion. In this case, the natural signals also contain other components: system-specific "resonances" and their interferential contributions at lower frequencies, and chaotic "spike" components at higher frequencies [12].

These regular and chaotic components can be separated and the appropriate parameters found using a general phenomenological approach to the analysis of complex stochastic signals $V(t)$: Flicker-Noise Spectroscopy (FNS) [12-16]. The key ideas of FNS are to treat different irregularities, such as spikes, "jumps", and discontinuities in derivatives of different orders, on all levels of the spatiotemporal hierarchy of the system under study as information carriers and to determine the correlation links in sequences of the values recorded for studied dynamic variables. In FNS, information parameters are introduced and determined for different time or space frequency ranges.

This paper first presents the fundamentals of FNS and then discusses the examples demonstrating that anomalous diffusion takes place in various complex processes. It is worth noting that all the above-mentioned data on the occurrence of anomalous diffusion in various processes are based on the use of Eq. (1), which describes random fluctuations in a virtually unlimited medium. On the other hand, the examples presented below correspond to the cases when the fluctuation variations of a dynamic variable reach a steady state (standard deviation) after some interval $T_1$. As expression (1) can be used only at small time intervals $\tau \ll T_1$, the appropriate FNS expressions were derived for the general case.

## II. FUNDAMENTALS OF FLICKER-NOISE SPECTROSCOPY

The advances of past decades in various approaches to the "Science of Complexity", including the analysis of the structure and dynamics of nonlinear systems using the theory of deterministic chaos [17, 18], nonlinear time series analysis [19], fractal theory [20, 21], and theory of cellular automata, which gave birth to the concepts of Self-Organized Criticality [22, 23] and the Principle of Computational Equivalence [24], made it possible to understand the physical principles of the chaotic evolution of open complex systems and the structures formed by it.

It was shown that the chaotic dynamics of open complex systems is associated with intermittency, consecutive alternation of rapid chaotic changes in the values of dynamic variables on small time intervals with small variations of the values on longer time intervals ("laminar" phases). Such intermittency occurs on every hierarchical level of the system evolution. It was shown that the origins of intermittency are associated with the occurrence of complex (multiparticle, nonlinear) interactions, dissipation, and inertia in these systems.

Therefore, intermittency leads to two different time scales in the chaotic dynamics of open complex systems for each hierarchical level: the higher-frequency one, corresponding to rapid changes, and the lower-frequency one, corresponding to the actual laminar phases. In this case the "energy" contribution of the rapid fluctuations may match or even exceed the corresponding contribution of the laminar phases to the power spectrum. One of the key FNS principles is to consider such intermittent evolution by assigning information parameters to different frequency ranges.

According to the FNS approach, the individual features of the evolution of complex systems are mostly contained in low-frequency (regular) components of the signals, i.e., internal and external resonances and their interferential contributions, which take place in the background of chaotic ("noise") high-frequency components associated with "spike" and "jump" irregularities of dynamic variable $V(t)$. The sequences of "spike" and "jump" irregularities reflect the intermittent nature of chaotic evolution with "spikes" corresponding to rapid fluctuations between laminar phases and "jumps" corresponding to "random walks", random stepwise changes in dynamic variable $V(t)$ in the region of laminar phases.

In FNS, all the introduced information is related to one of the fundamental concepts of statistical physics, the autocorrelation function

$$\psi(\tau) = \langle V(t)V(t+\tau) \rangle, \tag{2}$$

where $\tau$ is the time lag parameter. The angular brackets in relation (2) stand for the averaging over time interval $T$:

$$\langle (...) \rangle = \frac{1}{T} \int_{-T/2}^{T/2} (...) dt. \tag{3}$$

Expression (2) implies that function $\psi(\tau)$ describes the correlation between the values of dynamic variable $V(t)$ at smaller and larger values of the argument (assuming $\tau > 0$). The averaging over interval $T$ implies that all the characteristics that can be extracted by analyzing the $\psi(\tau)$ functions should be regarded as the average values for this interval. If the interval $T$ is a section of the larger interval $T_{tot}$ ($T < T_{tot}$), the value of function $\psi(\tau)$ can depend on the position of interval $T$ within the larger interval $T_{tot}$. If there is no such dependence and $\psi(\tau)$ is a function only of the difference in the arguments of the dynamic variables involved in (3), the evolution process being analyzed is defined as stationary. In this case, $\psi(\tau) = \psi(-\tau)$.

Discrete versions of Eqs. (2) and (3) and other formulas for experimental functions are presented in Appendix A of Ref. [13]. Continuous notation will be used throughout this paper for the reasons of compactness and ease of understanding.

To extract the information contained in $\psi(\tau)$ ($\langle V(t) \rangle = 0$ is assumed), one should analyze the transforms, or "projections", of this function, specifically the cosine transform («power spectrum» function) $S(f)$, where $f$ is the frequency:

$$S(f) = \int_{-T/2}^{T/2} \langle V(t)V(t+t_1) \rangle \cos(2\pi f t_1) \, dt_1 \qquad (4)$$

and its difference moments (Kolmogorov transient structural functions) of the second order $\Phi^{(2)}(\tau)$:

$$\Phi^{(2)}(\tau) = \langle [V(t) - V(t+\tau)]^2 \rangle. \qquad (5)$$

It is obvious that for the stationary process we have

$$\Phi^{(2)}(\tau) = 2[\psi(0) - \psi(\tau)], \qquad (6)$$

implying that $\Phi^{(2)}(\tau)$ depends linearly on $\psi(\tau)$.

Here, we use the cosine transform of the autocorrelation function (Fourier transform of a real signal is equal to the cosine transform for that signal) in contrast to the traditional estimate for power spectral density given as a square of the magnitude of the Fourier transform for the signal. The cosine-based estimate of power spectral density makes it possible to change from power spectral density to difference moment and vice versa by applying direct transformations to the autocorrelation function, which is used in section V to derive the procedure for separating regular and chaotic components from a stochastic signal. The Wiener-Khinchin theorem stating that the power spectral density is equal to the Fourier transform of the corresponding autocorrelation function is strictly formulated only for wide-sense stationary signals. The questions associated with applying this theorem to the extraction of information from real nonstationary signals measured on finite time intervals are discussed in section V.

As both $S(f)$ and $\Phi^{(2)}(\tau)$ are defined in terms of the autocorrelation function, one may assume that these functions and the parameters characterizing them are tightly interrelated. However, as will be shown below, the information contents of $S(f)$ and $\Phi^{(2)}(\tau)$ are different, and the parameters for both functions are needed to solve specific problems.

For simplicity, in this paper we will consider the problem of information difference between functions $S(f)$ and $\Phi^{(2)}(\tau)$ at a conceptual level. A more rigorous and substantiated analysis based on the theory of generalized functions [25] is presented elsewhere [12, 14]. Consider the process of one-dimensional "random walk" with small "kinematic viscosity" $V$ (Fig. 1).

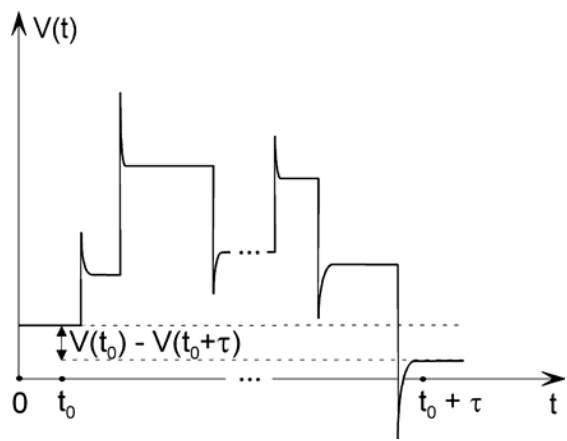

FIG. 1. One-dimensional "random walk" with small "kinematic viscosity".

The small value of $v$ implies that when the signal passes from position $V_i$ to $V_{i+1}$, which are $|V_{i+1} - V_i|$ apart (in value) from each other, the system first overleaps ("overreacts") due to inertia and then "relaxes". We assume that the relaxation time is small compared to the residence time in the "fluctuation" position. It is obvious that when the number of walks is large, the functions $\Phi^{(2)}(\tau)$ will be independent of the values of "inertial overleaps" of the system and depend only on the algebraic sum of walk "jumps". At the same time, the functions $S(f)$, which characterize the "energy side" of the process, will depend on both spikes and jumps.

In this regard, let us note a well-known result presented in chapter 4.3 of Schuster [17] (see Fig. 51), where an intermittent chaotic signal with alternating rapid chaotic spikes and laminar phases was considered. An artificial signal was generated by introducing a sequence of Dirac delta functions instead of rapid chaotic spikes. Then the power spectral density $S(f)$ for a sequence of $\delta$-functions with characteristic time intervals $T_0^i$ between adjacent $\delta$-functions on macroscopic time intervals $[-T/2, +T/2]$ ($T_0^i \ll T$) was calculated. It was shown that this artificial signal formed a flicker-noise dependency $S(f) \sim f^{-n}$ ($n \sim 1$) in the low-frequency spectrum range ($f \ll 1/2\pi T_0^i$). In other words, it was informative. On the other hand, if one would calculate the difference moment $\Phi^{(p)}(\tau)$ for this artificial signal (difference moment was not considered in [17]), it would be clear that it is equal to zero because the domain set of a $\delta$-function sequence is a set of measure zero [25]. It is easy to numerically illustrate this statement by replacing $\delta$-functions in calculating $\Phi^{(p)}(\tau)$ with one of the well-known approximations; for example, Gaussian with dispersion $\sigma_G^2$, and then passing to the limit $\sigma_G \to 0$.

It should be underlined that such separation of information stored in various irregularities is attributed to the intermittent character of the evolution dynamics. Indeed, the information contents of $S_c(f)$ and $\Phi_c^{(2)}(\tau)$ coincide if there is no intermittence, as shown for the case of completely "irregular" dynamics of the Weierstrass – Mandelbrot (WM) function in Appendix A. Here, we use the lower indices

"c" for functions $S_c(f)$ and $\Phi_c^{(2)}(\tau)$ to indicate the purely "chaotic" (no regular components) nature of the WM function.

The basic principles of FNS can be summarized as follows:

1. A hierarchy of spatiotemporal levels in complex open dissipative systems whose chaotic evolution is described by measured dynamic variable $V(t)$ on time interval $T$ is introduced.

2. The main information contained in chaotic signals $V(t)$ is provided by sequences of different types of irregularities such as spikes and jumps in the original signals and discontinuities in their derivatives of different orders on all levels of the spatiotemporal hierarchy of the system under study. The functions $\Phi_c^{(p)}(\tau)$ are formed exclusively by "jumps" of the dynamic variable while $S_c(f)$ is formed by both "spikes" and "jumps" on every level of the hierarchy.

3. The corresponding parameters for irregularities such as "discontinuities in derivatives" are extracted from the power spectra and difference moments built using time series $\Delta^m V(t_k)/\Delta t^m$ $(m \geq 1)$. Here, $\Delta^m V(t_k) = \Delta^{m-1} V(t_k) - \Delta^{m-1} V(t_{k-1})$; $\Delta t = t_k - t_{k-1}$ is the sampling interval for the values of the dynamic variable recorded at discrete times $t_k$.

4. Stationary processes in open dissipative systems, when the autocorrelator $\psi(\tau) = \langle V(t)V(t+\tau)\rangle$ depends only on the difference of arguments $\tau$, are characterized by a multi-parameter self-similarity, in contrast to the single-parameter self-similarity in the fractal and renormgroup theories. This implies that each introduced parameter has the same value for every spatiotemporal hierarchy level of the system.

### III. FNS EXPRESSIONS

Let us write the basic interpolation expressions for chaotic components that are used in the analysis of experimental time series, which were derived using the theory of generalized functions in Ref. [14]. The parameters characterizing the dynamic correlations on every level of the evolution hierarchy are assumed to be the same. Consider the simplest case, in which there is only one characteristic scale in the sequences of spikes and jumps [14]:

$$\Phi_c^{(2)}(\tau) \approx 2\sigma^2 \cdot \left[1 - \Gamma^{-1}(H_1) \cdot \Gamma(H_1, \tau/T_1)\right]^2, \tag{7}$$

$$\Gamma(s,x) = \int_x^\infty \exp(-t) \cdot t^{s-1} dt, \quad \Gamma(s) = \Gamma(s,0) \; .$$

Here, the lower index $c$ denotes the chaotic part of the signal (there are no regular components); $\Gamma(s)$ and $\Gamma(s,x)$ are the complete and incomplete gamma functions ($x \geq 0$ and $s > 0$), respectively; $\sigma$ is the standard deviation of the measured dynamic variable with dimension [$V$]; $H_1$ is the Hurst constant, which describes the rate at which the dynamic variable "forgets" its values on the time intervals that are less than the correlation time $T_1$. In this case, $T_1$ may be interpreted as the correlation time for the jumps in the chaotically varying time series $V(t)$.

For asymptotic cases, we obtain the formulas [14]:

$$\Phi_c^{(2)}(\tau) = 2\Gamma^{-2}(1+H_1) \cdot \sigma^2 \left(\frac{\tau}{T_1}\right)^{2H_1}, \quad \text{if } \frac{\tau}{T_1} \ll 1 \; ; \tag{8}$$

$$\Phi_c^{(2)}(\tau) = 2\sigma^2 \left[1 - \Gamma^{-1}(H_1) \cdot \left(\frac{\tau}{T_1}\right)^{H_1-1} \exp\left(-\frac{\tau}{T_1}\right)\right]^2, \quad \text{if } \frac{\tau}{T_1} \gg 1. \tag{9}$$

The interpolating function for power spectrum component $S_{cS}(f)$ formed by spikes can be written as [14]:

$$S_{cS}(f) \approx \frac{S_{cS}(0)}{1 + (2\pi f T_{0S})^{n_0}} \tag{10}$$

Here, $S_{cS}(0)$ is the parameter characterizing the low-frequency limit of $S_{cS}(f)$ and $n_0$ describes the degree of correlation loss in the sequence of spikes on the time interval $T_0$.

The interpolating function for the power spectrum component $S_{cR}(f)$ formed by jumps is written as [14]:

$$S_{cJ}(f) \approx \frac{S_{cJ}(0)}{1+(2\pi fT_1)^{2H_1+1}}, \tag{11}$$

where $S_{cJ}(0) = 4\sigma^2 T_1 H_1 \cdot \left\{1 - \frac{1}{2H_1 \Gamma^2(H_1)} \int_0^\infty \Gamma^2(H_1, \xi) d\xi\right\}$ is the parameter characterizing the low-frequency limit of $S_{cJ}(f)$ and $T_1$ is the corresponding correlation time.

The interpolating function for the overall power spectrum can be expressed as:

$$S_c(f) \approx \frac{S_c(0)}{1+(2\pi fT_0)^n}, \tag{12}$$

where $S_c(0)$, $T_0$, and $n$ are phenomenological parameters calculated by fitting the interpolating function (12) to the power spectrum for experimental time series. As FNS introduces the dynamic multiparametric self-similarity on all levels of the hierarchy, all the phenomenological parameters, including $H_1$ and $n$, are determined by fitting the interpolations (7) and (12) to the difference moment and power spectrum calculated using the experimental series. In other words, FNS does not impose any restrictions on the values of $H_1$ and $n$, as is the case of fractal theory: $0 \leq H_1 \leq 1$ and $n \leq 3$.

Although the contributions to the overall power spectrum $S_c(f)$ given by (10) and (11) are similar, the parameters in these equations are different: $S_{cS}(0) \neq S_{cJ}(0)$, $T_1 \neq T_{0S}$, and $2H_1 + 1 \neq n_0$. This implies that the parameters in the expressions for the power spectrum and structural function of the second order generally have different information contents when the experimental time series $V(t)$ is analyzed. For example, the characteristic times $T_{0S}$ and $T_1$ are usually much different because they correspond to different frequency bands of the power spectrum. The fact that jumps are more regular than spikes implies that the contribution of jumps to the power spectrum will be concentrated in its lower frequency band. At the same time, its higher frequency band, which is often characterized by the flicker-noise function $S_c(f) \sim 1/f^n$, is generated mostly by spikes.

## IV. RELATION BETWEEN DIFFUSION EQUATION AND FNS EXPRESSIONS

FNS is a phenomenological approach. The parameters introduced in FNS have a certain physical meaning and are determined by comparing the results calculated by FNS relations (7) – (12) with the curves calculated by Eqs. (4) and (5) using the experimental values of time series $V(t)$. For a stationary process, in which the autocorrelator $\psi(\tau) = \langle V(t)V(t+\tau) \rangle$ depends only on the difference in the arguments of dynamic variables and it is assumed that the ergodicity condition is met, the procedure of averaging over time (3) introduced in FNS is equivalent to the averaging procedure using the probability density function $W(V, t)$ for finding the values of the dynamic variable in the interval from $V$ to $V + dV$ at time $t$. In this case, expression (7) can be regarded as the generalized expression for the mean-squared deviation from the average value in the random walk processes described by Fickian equation or the equations of anomalous diffusion.

To find the relation between the phenomenological FNS parameters $\sigma$, $T_1$ and $H_1$ and the parameters characterizing the diffusion dynamics for a stationary process, we will first consider the simplest case of Fickian diffusion, for which $H_1 = 0.5$ Assume that the behavior of the probability density $W(V, \tau)$ (we use $\tau$ rather than $t$ as the process is stationary, with $\tau = 0$ being the start time) for the random variable $V$ on the segment $[-L, +L]$ over time $\tau$ can be described by the diffusion equation:

$$\frac{\partial W}{\partial \tau} = D \frac{\partial^2 W}{\partial V^2} \tag{13}$$

with the reflection (symmetry) conditions at the end points of the segment

$$\frac{\partial W}{\partial V} = 0 \text{ at } V = -L \text{ and } V = +L \tag{14}$$

and the initial condition

$$W(V, 0) = \delta(V). \tag{15}$$

Writing the Dirac delta function $\delta(V)$ as a series [26]:

$$\delta(V) = \frac{1}{2L}\left[1 + 2\sum_{k=1}^{\infty}\cos\frac{\pi k V}{L}\right],$$

we can find the solution to Eq. (13) subject to the above initial and boundary conditions:

$$W(V,\tau) = \frac{1}{2L}\left[1 + 2\sum_{k=1}^{\infty}\exp\left(-\frac{\pi^2 k^2 D\tau}{L^2}\right)\cos\frac{\pi k V}{L}\right]. \tag{16}$$

Now we can obtain the expressions for the average value of random variable $V$ and the mean-squared deviation of this variable from the average value after time $\tau$:

$$<V>_{pdf} = \frac{\int_{-L}^{+L} V W(V,\tau) dV}{\int_{-L}^{+L} W(V,\tau) dV} = 0, \tag{17}$$

$$\langle V^2 \rangle_{pdf} = \frac{\int_{-L}^{+L} V^2 W(V,\tau) dV}{\int_{-L}^{+L} W(V,\tau) dV} = \frac{4L^2}{\pi^2}\sum_{k=1}^{\infty}\frac{(-1)^{k+1}}{k^2}\left[1 - \exp\left(-\frac{\pi^2 k^2 D\tau}{L^2}\right)\right]. \tag{18}$$

From (18), we can derive the asymptotic expressions:

$$\langle V^2 \rangle_{pdf} \to 2D\tau \qquad \text{when } \tau << \frac{L^2}{\pi^2 D}, \tag{19}$$

$$\langle V^2 \rangle_{pdf} \to \frac{L^2}{3} \qquad \text{when } \tau >> \frac{L^2}{\pi^2 D}. \tag{20}$$

Asymptotic expression (20) was found using the formula [27]:

$$\sum_{k=1}^{\infty}\frac{(-1)^{k+1}}{k^2} = \frac{\pi^2}{12},$$

Expression (19) was found using the first derivative of Eq. (16) with respect to time $\tau$:

$$\frac{d\langle V^2 \rangle_{pdf}}{d\tau} = 4D\sum_{k=1}^{\infty}(-1)^{k+1}\exp(-\xi k^2), \tag{21}$$

where $\xi = \pi^2 D\tau / L^2 << 1$:

$$\sum_{k=1}^{\infty}(-1)^{k+1}\exp(-\xi k^2) = \sum_{n=1}^{\infty} \Delta n\left\{\exp\left[-4\xi\left(n-\frac{1}{2}\right)^2\right] - \exp(-4\xi n^2)\right\} =$$

$$= \frac{1}{2\xi}\left[\int_0^{2\xi}\exp\left(-\frac{x^2}{\xi}\right)dx - \int_0^{\xi}\exp\left(-\frac{x^2}{\xi}\right)dx\right] = \frac{1}{4}\left(\frac{\pi}{\xi}\right)^{1/2}\left[\Phi\left(2\xi^{1/2}\right) - \Phi\left(\xi^{1/2}\right)\right] \xrightarrow[\xi\to 0]{} \frac{1}{2}, \qquad (22)$$

where $\Phi(x)$ is the error integral [27]. In Eq. (22), the increment in the discrete values of $n$ in the summation was formally taken to be $\Delta n = 1$, followed by the transition from summation to integration using the integration variable $x = \xi n$ and the differential $d\xi = \xi \Delta n \ll 1$.

The relation between the parameters of the diffusion problem and phenomenological FNS parameters can be found by comparing asymptotic expressions (19) and (20) for small and large values of $\tau$ with the corresponding expressions (8) and (9) written for $H_1 = 0.5$:

$$D = \frac{4}{\pi} \cdot \frac{\sigma^2}{T_1}; \quad L^2 = 6\sigma^2. \qquad (23)$$

In this case, the difference between the values calculated by expressions (18) and (7) with $H_1 = 0.5$ for the range of intermediate values of parameter $\tau$ does not exceed 20% (curve 1 in Fig. 2). The figure demonstrates the normalized curves calculated by Eqs. (7) and (18), for which the asymptotic values at $\tau/T_1 \ll 1$ and $\tau/T_1 \gg 1$ coincide:

$$\phi_1(\tau) \equiv \frac{3}{L^2}\langle V^2\rangle_{pdf}, \quad \phi_2(\tau) \equiv \frac{1}{2\sigma^2}\Phi^{(2)}(\tau). \qquad (24)$$

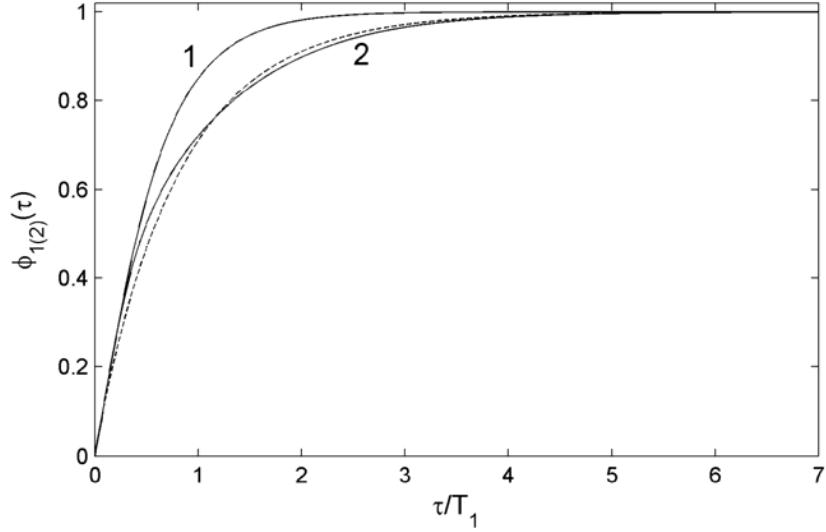

FIG. 2. Normalized functions $\phi_1(\tau)$ (curve 1 for boundary conditions (14), curve 2 for boundary conditions (29)-(30)) and $\phi_2(\tau)$ (dashed).

As the analysis of a large number of natural signals and the examples presented below demonstrate, the phenomenological expression (7) describes the chaotic component of experimental structural functions much better than the model expression (18). This is attributed to the limitations of boundary conditions (14) used in deriving Eq. (18) with regard to real systems. In order to solve the diffusion problem with a steady-state limit at $\tau \to \infty$, it would be more accurate to select not fixed boundaries [– L, + L], but varying boundaries, and use integral reflection conditions. To this end, generalized boundary conditions [28], which take into consideration the effects of nonstationarity and the finite residence times of the diffusion system in the "adstates" of boundaries + L and – L, can be used. Introducing the probability densities $w_{\pm L}(\tau)$ for finding the system in such boundary "adstates", the modified boundary conditions can be written as:

At $V = -L$:

$$D\frac{\partial W}{\partial V} = \chi W(-L, \tau) - \lambda w_{-L}(\tau), \qquad (25)$$

$$\frac{dw_{-L}(\tau)}{dt} = \chi W(-L,\tau) - \lambda w_{-L}(\tau). \tag{26}$$

At $V = +L$:

$$-D\frac{\partial W}{\partial V} = \chi W(+L,\tau) - \lambda w_{+L}(\tau), \tag{27}$$

$$\frac{dw_{+L}(t)}{dt} = \chi W(+L,\tau) - \lambda w_{+L}(\tau). \tag{28}$$

Here, $\chi$ and $\lambda$ are the rate constants for direct and reverse transitions of the system from a boundary "diffusion" state to an "adstate", respectively. Assume that their values are the same for both boundaries.

After solving Eqs. (26) and (28) for probability densities $w_{-L}(\tau)$ and $w_{+L}(\tau)$ and substituting the solutions into Eqs. (25) and (27), the boundary conditions take the following form:

$$D\frac{\partial W}{\partial V} = \chi W(-L,\tau) - \lambda \chi \exp(-\lambda\tau)\int_0^\tau W(-L,\xi)\exp(\lambda\xi)d\xi \quad \text{at} \quad V = -L, \tag{29}$$

$$-D\frac{\partial W}{\partial V} = \chi W(L,\tau) - \lambda \chi \exp(-\lambda\tau)\int_0^\tau W(L,\xi)\exp(\lambda\xi)d\xi \quad \text{at} \quad V = -L. \tag{30}$$

The integrodifferential problem given by Eqs. (13), (15), (29)-(30) was numerically solved using the iterative procedure described in Appendix B. The following values of parameters were used: $\chi = 0.4$, $\lambda = 0.04$. As can be seen in Fig. 3, the curve calculated for boundary conditions (29)-(30) approaches much closer to the interpolation formula (7) than the curve for the boundary conditions of symmetry. The relative error for intermediate values of $\tau$ in this case does not exceed 5 %. In other words, interpolation (7) is practically equivalent to the mean-squared deviation calculated by solving the problem (13), (15), (29)-(30), i.e., diffusion model with integrodifferential boundary conditions, which can be used for simulating the chaotic processes in the real complex systems with anomalous diffusion.

It is difficult to compare the functions $<V^2>_{pdf}$ and $\Phi^{(2)}(\tau)$ for anomalous diffusion ($H_1 \neq 0.5$) because the corresponding equations for the probability density $W(V, \tau)$ of random variable $V$ varying on the segment $[-L, +L]$ over time $\tau$ are very complicated and can be solved only by numerical methods [3, 4, 9]. At the same time, the desired relation can be found if we compare expressions (8) and (1), which correspond to the case of anomalous diffusion at small values of $\tau$, by choosing $T_1$ as the characteristic time $t_0$, and assume that in this case the FNS parameter $\sigma$ corresponds to some model parameter $L_a$ determining the region in the range of values of the dynamic variable where the states of the system can be localized:

$$D = \frac{1}{\Gamma^2(1+H_1)} \cdot \frac{\sigma^2}{T_1}; \quad L_a^2 = b\sigma^2, \tag{31}$$

Here, $b$ is a dimensionless parameter.

## V. SEPARATION OF REGULAR COMPONENTS FROM A SIGNAL

As was noted above, the stochastic dynamics of complex systems includes both chaotic components, i.e., "spike" and "jump" irregularities, and system-specific slowly varying regular components associated with a set of frequencies. These frequencies correspond to internal and external resonances and their interferences. It should be noted that the whole set of the resonance and interferential frequencies may get rearranged during the evolution of an open system. All the specific frequencies and their interferential contributions, which manifest themselves as oscillations in the dynamic variable $V(t)$, will be further called as "regular" components.

In the general case, the signal $V(t)$ under study can be formally written as

$$V(t) = V_{cS}(t) + [V(t) - V_{cS}(t)] = V_{cS}(t) + V_{cJ}(t) + V_r(t), \tag{32}$$

where

$$V_r(t) \equiv V(t) - V_{cS}(t) - V_{cJ}(t).$$

Here, $V_{cS}(t)$ and $V_{cJ}(t)$ are the chaotic components formed by spikes (mostly, the highest-frequency band) and jumps (mostly, the intermediate-frequency band), respectively; $V_r(t)$ is the low-frequency signal formed by regular components, which are characterized by a gradual variation against the background of mostly high-frequency chaotic components.

Let us assume that the chaotic components $V_{cS}(t)$ and $V_{cJ}(t)$ do not correlate with low-frequency components $V_r(t)$, i.e.:

$$\psi_{rcS}(\tau) \equiv \langle V_r(t)V_{cS}(t+\tau)\rangle = \langle V_{cS}(t)V_r(t+\tau)\rangle = 0, \\ \psi_{rcJ}(\tau) \equiv \langle V_r(t)V_{cJ}(t+\tau)\rangle = \langle V_{cJ}(t)V_r(t+\tau)\rangle = 0. \tag{33}$$

In addition, it is assumed that

$$\psi_r(\tau) = \langle V_r(t)V_r(t+\tau)\rangle = \psi_r(-\tau). \tag{34}$$

Relations (32)-(34) make it possible to characterize the signal $V(t)$ under study and determine the parameters of anomalous diffusion. First, it is necessary to subtract the interpolation function $S_c(f)$ [expression (12)], which corresponds to the contribution of high-frequency components $V_{cS}(t)$ and $V_{cJ}(t)$ to $S(f)$, from the spectrum $S(f)$, determine $S_c(0)$ as the minimum power spectrum value for first several frequency points, and use a trust-region method of nonlinear least squares, built into MathWorks MATLAB v. 7.1, to determine the parameters $n$ and $T_0$. It may be assumed that $n = n_0$ in the region of highest frequencies. The final conclusion about the validity of this equality, the determination of $T_0$, and the estimation of each component in the sum $S_c(0) = S_{cJ}(0) + S_{cS}(0)$ is performed later, when the parameters $\sigma$, $T_1$, and $H_1$ are known.

Using the difference

$$S_r(f) = S(f) - S_c(f), \tag{35}$$

calculate the "incomplete reverse cosine-transform":

$$\varphi_r(\tau) = 2\int_0^{f_{max}} S_r(f)\cos(2\pi f\tau)df, \tag{36}$$

where $f_{max} = \dfrac{f_s}{2} = \dfrac{\Delta t^{-1}}{2} = \dfrac{N}{2T}$; $f_s$ is the sampling frequency; $\Delta t$ is the time interval between the adjacent sampled values of signal $V(t)$, the total number of which is equal to $N$.

Next, the total contribution made by the regular component to the difference moment of the second order $\Phi^{(2)}(\tau)$ is calculated based on Eqs. (36) and (6):

$$\Phi_r^{(2)}(\tau) = 2[\varphi_r(0) - \varphi_r(\tau)], \qquad (37)$$

After that the contribution made by the chaotic component to the overall function $\Phi^{(2)}(\tau)$ is estimated as

$$\Phi_c^{(2)}(\tau) = \Phi^{(2)}(\tau) - \Phi_r^{(2)}(\tau). \qquad (38)$$

The comparison of experimental data and the values calculated by relations (36)-(38) using a trust-region method of nonlinear least squares, built into MathWorks MATLAB v. 7.1, is performed to determine the values of parameters $\sigma$, $T_1$, and $H_1$, which characterize the contribution made by jumps into the structural function. In this case, the values of parameters $H_1$, $\sigma$, and $T_1$ are chosen by providing the best agreement between the experimental and calculated curves $\Phi^{(2)}(\tau)$ in the entire interval of $\tau$ under study. As a result, the above parameter $H_1$ is somewhat different in meaning from the Hurst constant, which is usually introduced for describing the functions $\Phi^{(2)}(\tau)$ at small values of $\tau$.

Detailed description of the FNS parameterization procedure with the appropriate expressions in the discrete format is presented in Appendix C.

As real signals are nonstationary and recorded on finite intervals, it is necessary to discuss how the interpolation expressions (7) and (10)-(12) for the chaotic components of power spectrum estimates $S(f)$ and difference moments $\Phi^{(2)}(\tau)$ presented in section III and expression (37), which is used for calculating the regular component of $\Phi^{(2)}(\tau)$, can be used in this case. Expressions (7), (10)-(12) were derived in [6, 8] for an intermittent stationary signal $V(t)$ on a time interval $T$ with $T \to \infty$. According to the general view of intermittent evolution, two types of irregularities at two different frequency ranges were introduced: jump irregularities, represented as Heaviside theta functions, and spike irregularities, represented as Dirac delta functions. Using the theory of generalized functions, the integral expressions for $S(f)$ and $\Phi^{(2)}(\tau)$ were derived in the low-frequency limit, when the considered time intervals greatly exceed the intervals between adjacent irregularities. The kernels of these functions, which characterize the correlation links in sequences of irregularities, were built based on simple phenomenological expressions with corresponding parameters. As expressions (7), (10)-(12), and (35)-(38) were derived based on phenomenological kernels, the expressions themselves should also be considered phenomenological, and thus can be used for nonstationary signals. The subsequent analysis [12, 13] and the results presented in section VI show that the expressions (7), (10)-(12) for the chaotic component and expressions (35)-(38) for the regular component adequately describe experimental data for various complex systems, which demonstrates that the phenomenological expressions for the kernels were chosen properly.

There are signals for which approximations (7) and (10)-(12) cannot adequately describe experimental variations. In this case, it is necessary to use more complex phenomenological expressions for the kernels of the integral expressions for $S(f)$ and $\Phi^{(2)}(\tau)$ with a larger number of empirical parameters (see expression (25) in Ref. [14] or Ref. [29] for more information). This problem will be discussed in detail in section VI.B.

## VI. EXAMPLES OF ANOMALOUS DIFFUSION IN COMPLEX SIGNALS

The purpose of this section is to show using the FNS parameterization procedure described above that anomalous diffusion can be seen in various complex processes.

## A. Anomalous diffusion in magnetoencephalograms

Consider the analysis of the dynamic characteristics of neuromagnetic cortex responses (magnetoencephalograms, MEG) to flickering-color stimuli RB (red-blue), which were studied in Refs. [30-34]. The experimental setup shown in Fig. 3 was used for collecting the data generated by the 61-SQUID (superconducting quantum interference device) sensors attached to different points around the head, which can record weak magnetic induction gradients of about $10^{-11}$-$10^{-10}$ T/cm. The sampling frequency $f_d$ of MEG signals was 500 Hz ($f_d = 500\ Hz$). The goal of the research is to study the potential danger of some modern cartoons to provoke photosensitive epilepsy (PSE) in children. The MEG signals of healthy subjects (from the control group of 9 volunteers) and a 12-year old PSE patient were recorded [30, 31].

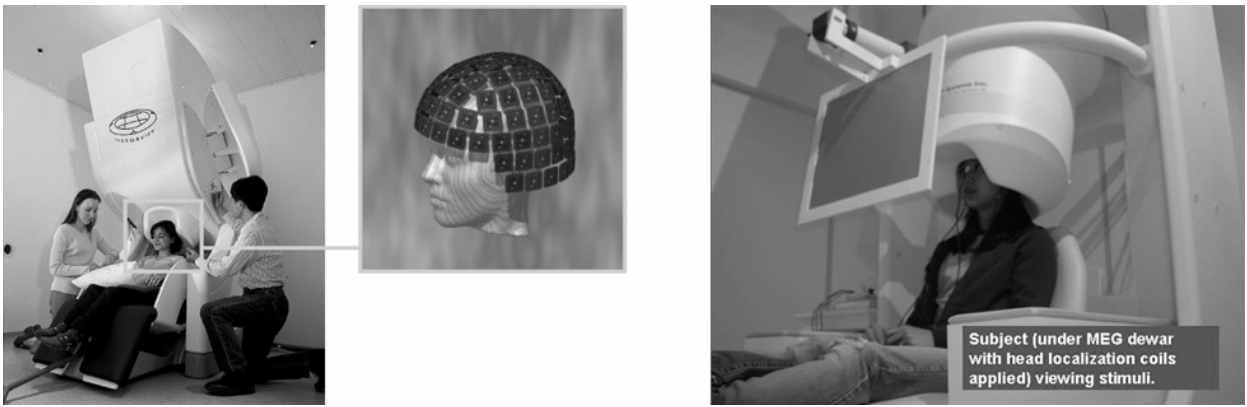

FIG. 3. Sample setup for recording MEG signals and the scheme for placing SQUID-sensors [34].

It was found earlier that among the sensors recording MEG response signals, sensor 10, which is located at the frontal lobe on the head, shows the highest sensitivity to these color stimuli [30, 31]. So, in this example we will consider the analysis of the MEG responses for two volunteers (7-th and 9-th) recorded from this sensor. For comparison, we will also show the results for sensor 43, which is located on the scalp over the occipital lobe of the cerebral cortex. The results for the patient [34] will not be listed here because they are characterized by a large number of high-frequency resonances, which would require a higher sampling frequency of the measurements to determine the parameters of anomalous diffusion. More complete analysis of the MEG signals is presented elsewhere [34].

Figures 4-7 illustrate the MEG signals recorded by sensors 10 and 43 as the response to the RB stimulus for healthy subjects 7 and 9. Every figure also demonstrates the results of the FNS analysis of the recorded signals, with the FNS parameters being given in the figure captions. It should be noted that the collections of the parameters are highly specific.

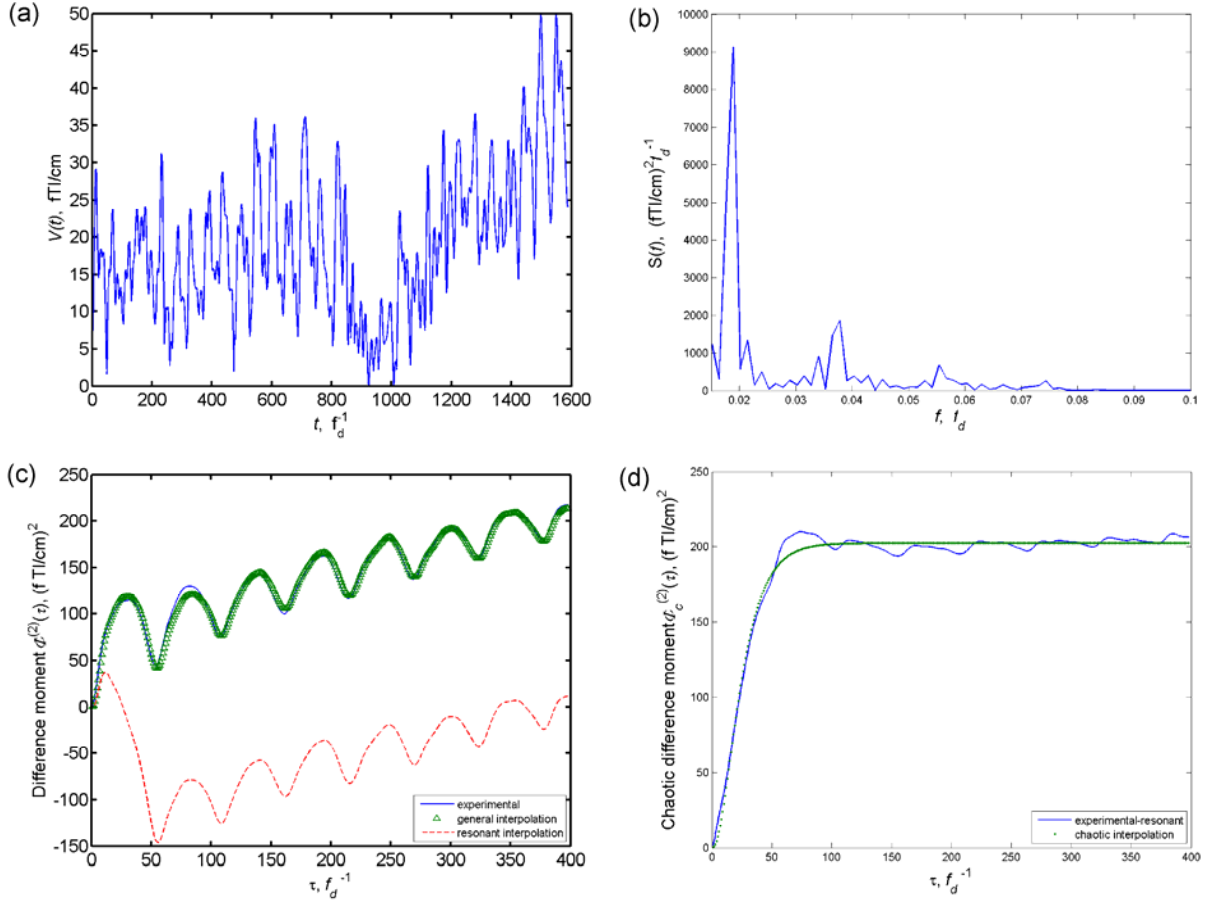

FIG. 4. Analysis of the MEG signal recorded at sensor 10 for control subject 7 as the response to RB-stimulus ($T = 3.2$ s; $\sigma =10.1$ fTl/cm, $H_1 = 1.27$, $T_1 = 2.9 \cdot 10^{-2}$ s, $D \approx 3.0 \cdot 10^3$ fTl$^2$/(cm$^2$ s), $S_c(0) = 1.07 \cdot 10^4$ fTl$^2$ / (cm$^2 f_d$), $T_0 = 3.8 \cdot 10^{-2}$ s, $n = 3.2$): (a) source signal [30-31]; (b) power spectrum $S(f)$ given by Eq. (4) in the low-frequency range (main peaks: 1.6 – 9.4 – 22.8 – 27.8 Hz); (c) "experimental" – $\Phi^{(2)}(\tau)$ given by Eq. (5), "general interpolation" – $\Phi^{(2)}(\tau)$ given by Eq. (38), "resonant interpolation" – $\Phi_r^{(2)}(\tau)$ given by Eq. (37); (d) – "experimental – resonant" – $\Phi^{(2)}(\tau)$ given by Eq. (38) minus $\Phi_r^{(2)}(\tau)$ given by Eq. (37), "chaotic interpolation" – $\Phi_c^{(2)}(\tau)$ given by Eq. (7).

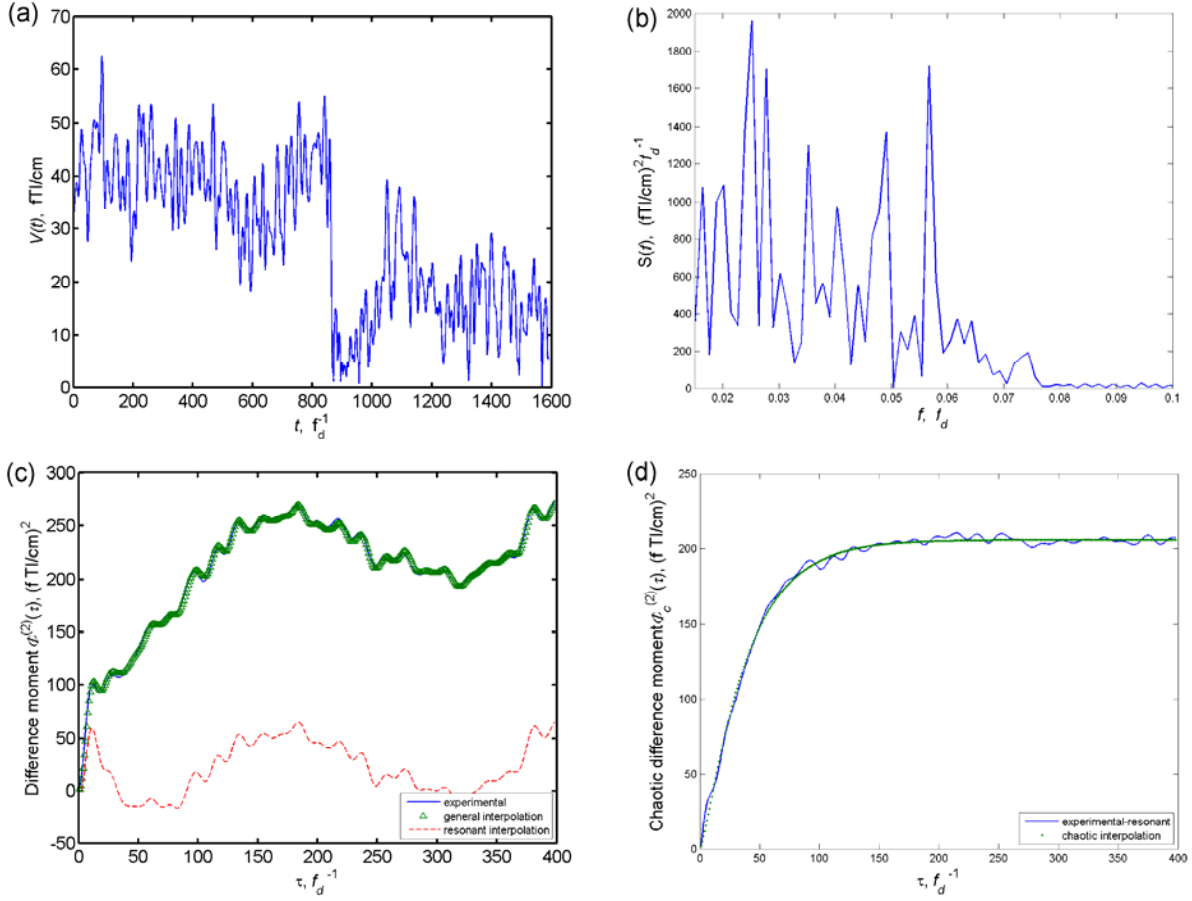

FIG. 5. Analysis of the MEG signal recorded at sensor 10 for control subject 9 as the response to RB-stimulus ($T = 3.2$ s; $\sigma = 10.1$ fTl/cm, $H_1 = 0.67$, $T_1 = 7.5 \cdot 10^{-2}$ s, $D \approx 1.50 \cdot 10^3$ fTl$^2$/(cm$^2$ s), $S_c(0) = 1.74 \cdot 10^4$ fTl$^2$ / (cm$^2$ $f_d$), $T_0 = 7.4 \cdot 10^{-2}$ s, $n = 2.2$): (a) source signal [30-31]; (b) power spectrum $S(f)$ given by Eq. (4) in the low-frequency range (main peaks: 1.7 – 6 – 12.5 – 24.5 – 28.5 Hz); (c) "experimental" – $\Phi^{(2)}(\tau)$ given by Eq. (5), "general interpolation" – $\Phi^{(2)}(\tau)$ given by Eq. (38), "resonant interpolation" – $\Phi_r^{(2)}(\tau)$ given by Eq. (37); (d) – "experimental – resonant" – $\Phi^{(2)}(\tau)$ given by Eq. (38) minus $\Phi_r^{(2)}(\tau)$ given by Eq. (37), "chaotic interpolation" – $\Phi_c^{(2)}(\tau)$ given by Eq. (7).

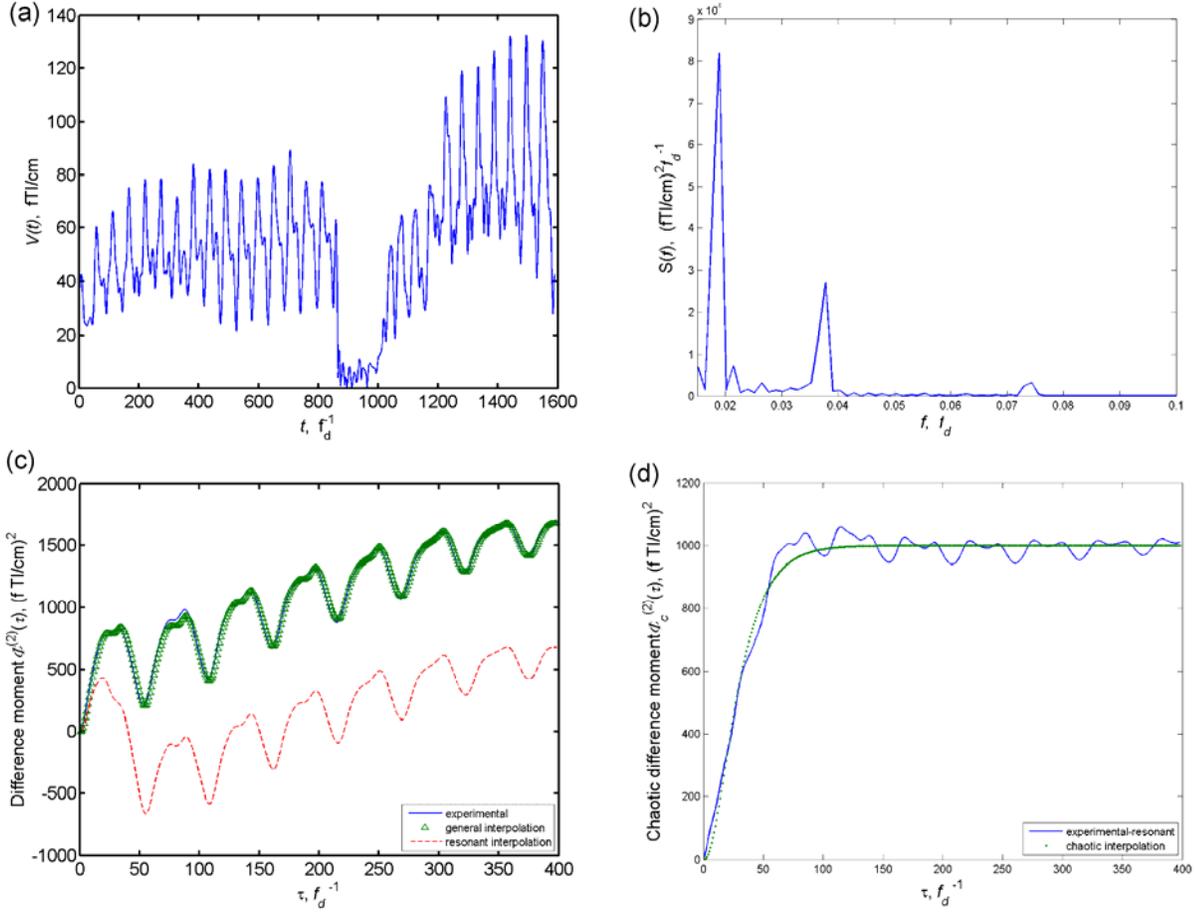

FIG. 6. Analysis of the MEG signal recorded at sensor 43 for control subject 7 as the response to RB-stimulus ($T = 3.2$ s; $\sigma = 22.4$ fTl/cm, $H_1 = 1.25$, $T_1 = 3.52 \cdot 10^{-2}$ s, $D \approx 1.2 \cdot 10^4$ fTl$^2$/(cm$^2$ s), $S_c(0) = 6.1 \cdot 10^4$ fTl$^2$ / (cm$^2$ $f_d$), $T_0 = 4.4 \cdot 10^{-2}$ s, $n = 3.3$): (a) source signal [30-31]; (b) power spectrum $S(f)$ given by Eq. (4) in the low-frequency range (main peaks: 1.6 – 12.6 – 19.1 – 37.5 Hz); (c) "experimental" – $\Phi^{(2)}(\tau)$ given by Eq. (5), "general interpolation" – $\Phi^{(2)}(\tau)$ given by Eq. (38), "resonant interpolation" – $\Phi_r^{(2)}(\tau)$ given by Eq. (37); (d) – "experimental – resonant" – $\Phi^{(2)}(\tau)$ given by Eq. (38) minus $\Phi_r^{(2)}(\tau)$ given by Eq. (37), "chaotic interpolation" – $\Phi_c^{(2)}(\tau)$ given by Eq. (7).

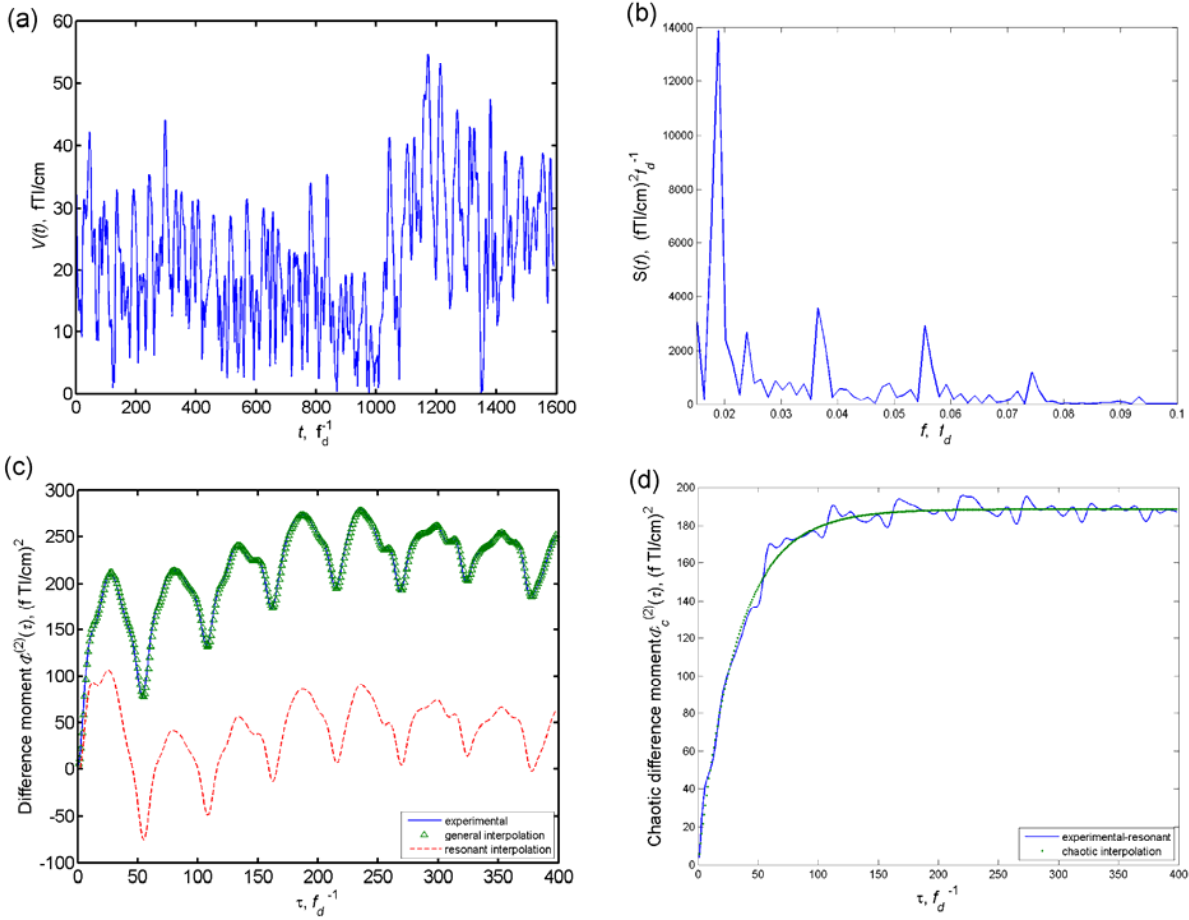

FIG. 7. Analysis of the MEG signal recorded at sensor 43 for control subject 9 as the response to RB-stimulus ($T = 3.2$ s; $\sigma = 9.7$ fTl/cm, $H_1 = 0.51$, $T_1 = 8 \cdot 10^{-2}$ s, $D \approx 1.3 \cdot 10^3$ fTl$^2$/(cm$^2$ s), $S_c(0) = 1.03 \cdot 10^4$ fTl$^2$ / (cm$^2$ $f_d$), $T_0 = 5.98 \cdot 10^{-2}$ s, $n = 2.2$): (a) source signal [30-31]; (b) power spectrum $S(f)$ given by Eq. (4) in the low-frequency range (main peaks: 2.5 – 10 – 17.5 – 28 – 37.5 Hz); (c) "experimental" – $\Phi^{(2)}(\tau)$ given by Eq. (5), "general interpolation" – $\Phi^{(2)}(\tau)$ given by Eq. (38), "resonant interpolation" – $\Phi_r^{(2)}(\tau)$ given by Eq. (37); (d) – "experimental – resonant" – $\Phi^{(2)}(\tau)$ given by Eq. (38) minus $\Phi_r^{(2)}(\tau)$ given by Eq. (37), "chaotic interpolation" – $\Phi_c^{(2)}(\tau)$ given by Eq. (7).

As can be seen from Figs. 4(d)-7(d), the chaotic parts of the difference moments for the signals, which are obtained by subtracting the regular components from the experimental difference moments, can be adequately described by the anomalous-diffusion interpolation (7). In other words, human brain neural activity contains the chaotic components that can be modeled using the equations of anomalous diffusion. It is worth noting that in addition to the chaotic components there are regular ones, which are extremely difficult to model because they get rearranged *in vivo*. It should be emphasized that the adequate description of the experimental data was achieved using expression (7) for the chaotic difference-moment component $\Phi_c^{(2)}(\tau)$, which is practically equivalent to the mean-squared-deviation function (24) obtained by the numerical solution of problem (13), (15), (29)-(30) (for the case when $H_1 \approx 0.5$, which corresponds to Fig. 7). On the other hand, the mean-squared-deviation function obtained by solving problem (13)-(15) could not adequately describe the experimental curve in Fig. 7, which implies that the symmetry boundary conditions given by (14) cannot be used for modeling real processes.

The above FNS analysis of MEG response signals demonstrates a principal possibility of separating the contribution of the anomalous-diffusion component out of the overall complex dynamics of studied biomedical processes, implying that this can be used to simulate the separate functioning stages of a cerebral cortex *in vivo*.

**B. Anomalous diffusion in blinking fluorescence of quantum dots**

The recent progress in nanotechnologies gave rise to several new problems related to the standardization and stabilization of the functional activity of quantum-sized objects. Particular examples are "quantum dots" (QD), which are regarded as the functional elements of the quantum computers of the future, and photoactive elements, which are added or formed in inorganic or organic matrices during the production of optical materials (photochromes, luminophores). Consider the problem of stabilizing the functional parameters of the elements using the example of blinking intermittent dynamics of the emission by a single quantum dot in a system of colloid quantum dots [35, 36].

We will analyze the fluorescence signal generated by a CdSe quantum dot overcoated with a thin layer of ZnS (so-called core-shell QD). The experimental data were kindly provided by Professor Masaro

Kuno (University of Notre Dame). The radius of the core CdSe particle was 2.7 nm. There were 3 "monolayers" of ZnS surrounding this particle. The laser excitation wavelength was 488 nm and the incident intensity was 600 W/cm$^2$. The absorption cross section of such dots was approximately $10^{-15}$ cm$^2$. The quantum yield of these dots at the ensemble level should be of the order of ~30%. The signal was recorded during $T_{tot}$ = 1 hour at a sampling frequency $f_d$ = 100 Hz. So, the time series contains 360,000 values in the time interval of $1 - 360000\, f_d^{-1}$. During this time interval the fluorescence intensity significantly dropped. To demonstrate the nonstationarity of the signal, which manifests itself in the dependence of the FNS parameters on the averaging window selected within the interval $T_{tot}$, in addition to the source signal we also analyzed its three slices with 36,000 values each, corresponding to the initial, intermediate, and final stages.

Figures 8(a)-11(a) show the source signal and its slices at intervals I ($1 - 36000\, f_d^{-1}$), II ($162001 - 198000\, f_d$) and III ($324001 - 360000\, f_d$), respectively. The graphs of the relations used in the FNS analysis are presented in Figs. 8(b)-(d) – 11(b)-(d), with the FNS parameters being given in the figure captions. To reduce the fitting error in describing the experimental difference moment, the zeroth frequency point was excluded from the regular difference moment component, i.e., $q_{\min} = 1$ was used in the parameterization algorithm listed in Appendix C. It can be seen that the chaotic part $\Phi_c^{(2)}(\tau)$ of experimental difference moments for all four signals is adequately approximated by the anomalous-diffusion interpolation (7), which in this case corresponds to "subdiffusion". The differences in the values of FNS parameters for the slices at intervals I, II, and III imply that the signal is nonstationary. Fifty fluorescence signals produced by other single QDs were also analyzed (not listed here). For some of the signals, the anomalous-diffusion interpolation (7) could not adequately describe the chaotic part of the experimental difference moment. In these cases, one should use more complex phenomenological interpolations than the 3-parameter expression (7); for example, the interpolation formula discussed in Ref. [37].

The above analysis gives an illustration of a nanoscale open complex system the stochastic dynamics of which is governed by anomalous diffusion.

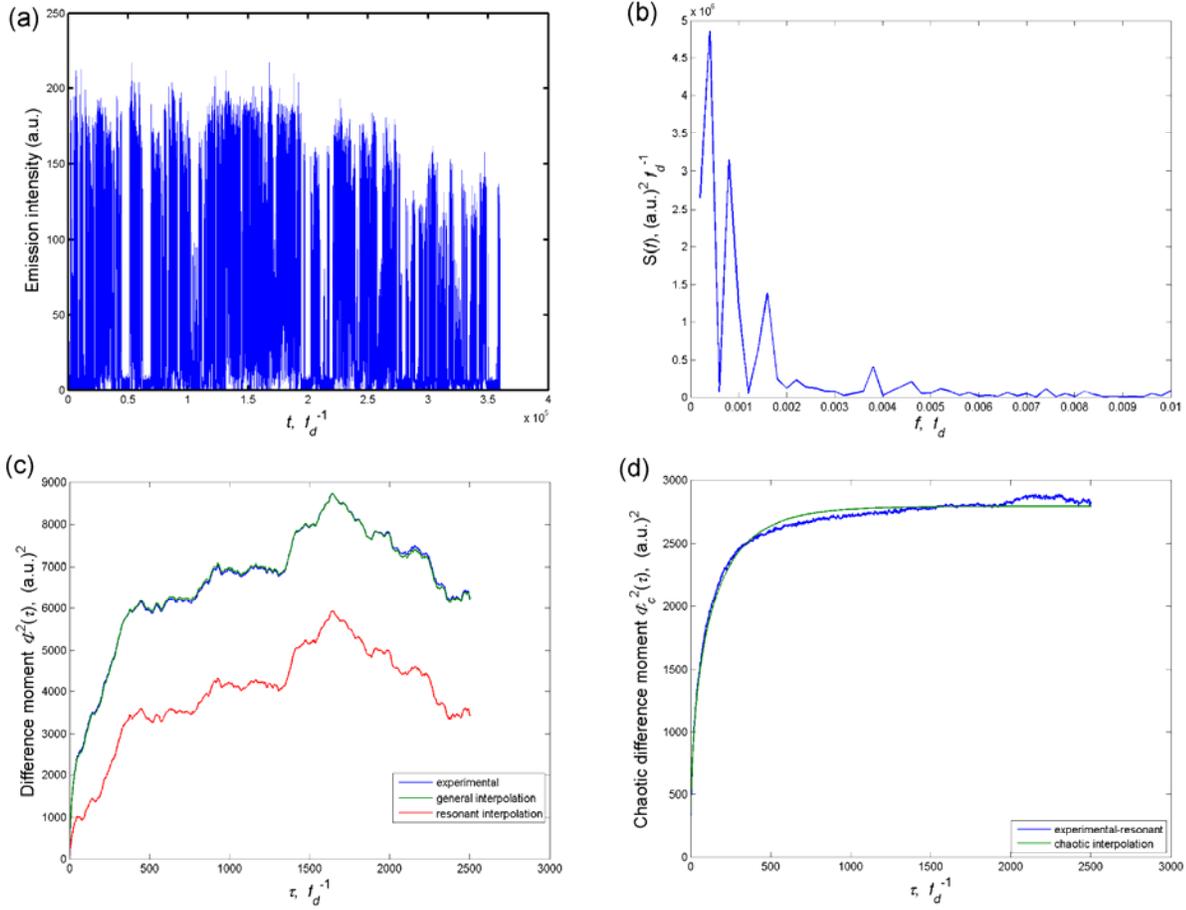

FIG. 8. Analysis of the fluorescence signal generated by a CdSe quantum dot overcoated with a thin layer of ZnS (so-called core-shell QD) on the time interval from 1 to 360000 $f_d^{-1}$ ($\sigma = 37.4$ arb. un.; $H_1 = 0.21$; $T_1 = 334\,\Delta t = 3.34$ s, $D \approx 453.4$ (arb. un.)$^2$/s, $S_c(0) = 5.88\cdot 10^5$ (arb. un.)$^2 f^{-1}_d$; $n = 1.43$; $T_0 = 175{,}5\,\Delta t \approx 1{,}75$ s): (a) source signal; (b) power spectrum $S(f)$ given by Eq. (4) in the low-frequency range (100 frequencies); (c) "experimental" – $\Phi^{(2)}(\tau)$ given by Eq. (5), "general interpolation" – $\Phi^{(2)}(\tau)$ given by Eq. (38), "resonant interpolation" – $\Phi_r^{(2)}(\tau)$ given by Eq. (37); (d) – "experimental – resonant" – $\Phi^{(2)}(\tau)$ given by Eq. (38) minus $\Phi_r^{(2)}(\tau)$ given by Eq. (37), "chaotic interpolation" – $\Phi_c^{(2)}(\tau)$ given by Eq. (7).

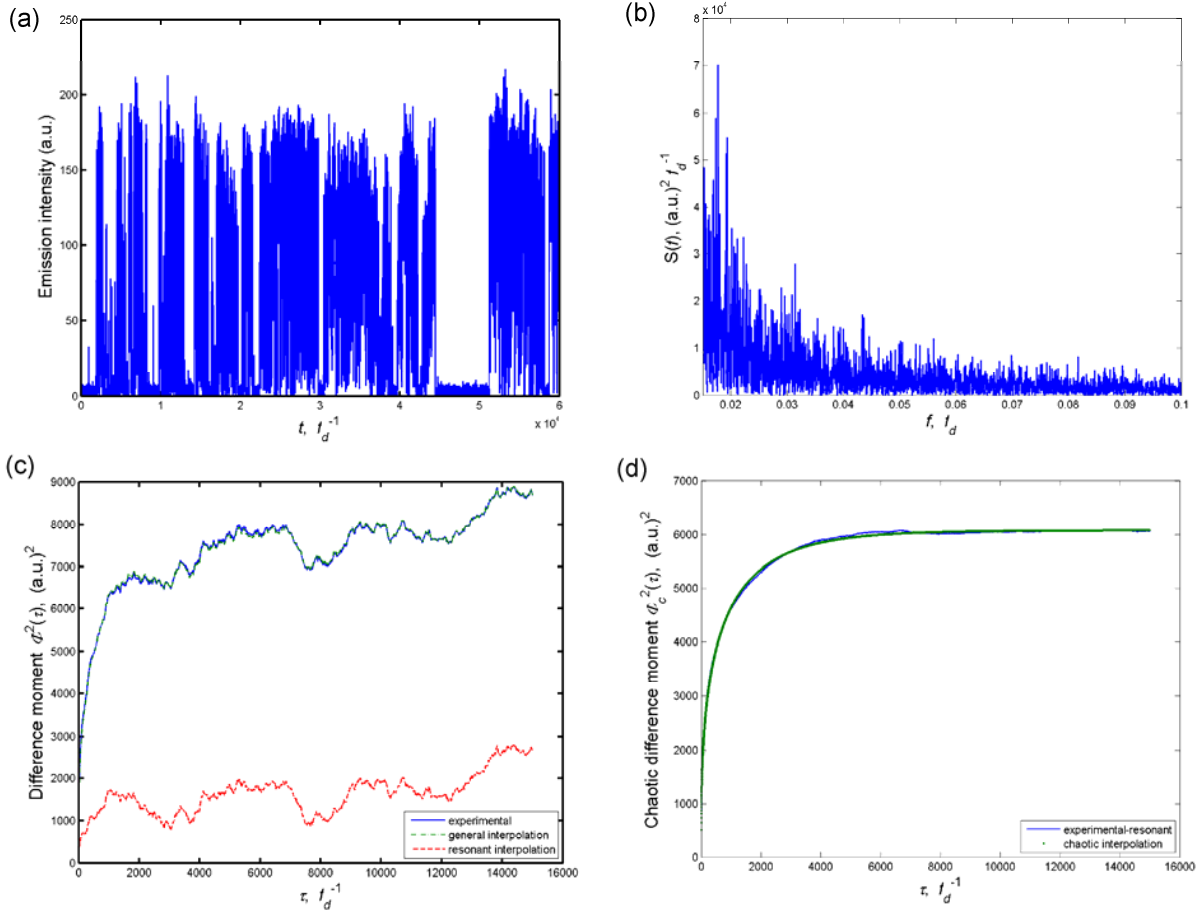

FIG. 9. Analysis the signal in Fig. 8 on the time interval from 1 to 36000 $f_d^{-1}$ ($\sigma$ = 55.2 arb. un.; $H_1$ = 0.17; $T_1$ = 2598.6 $\Delta t \approx$ 26 s relative to the interval start, $D \approx$ 125.6 (arb. un.)$^2$/s, $S_c(0)$ = 1.71·10$^6$ (arb. un.)$^2 f^{-1}_d$; $n$ = 1.25; $T_0$ = 3247.4 $\Delta t \approx$ 32.5 s relative to the interval start; $q_{min}$ =1 - see Appendix C).

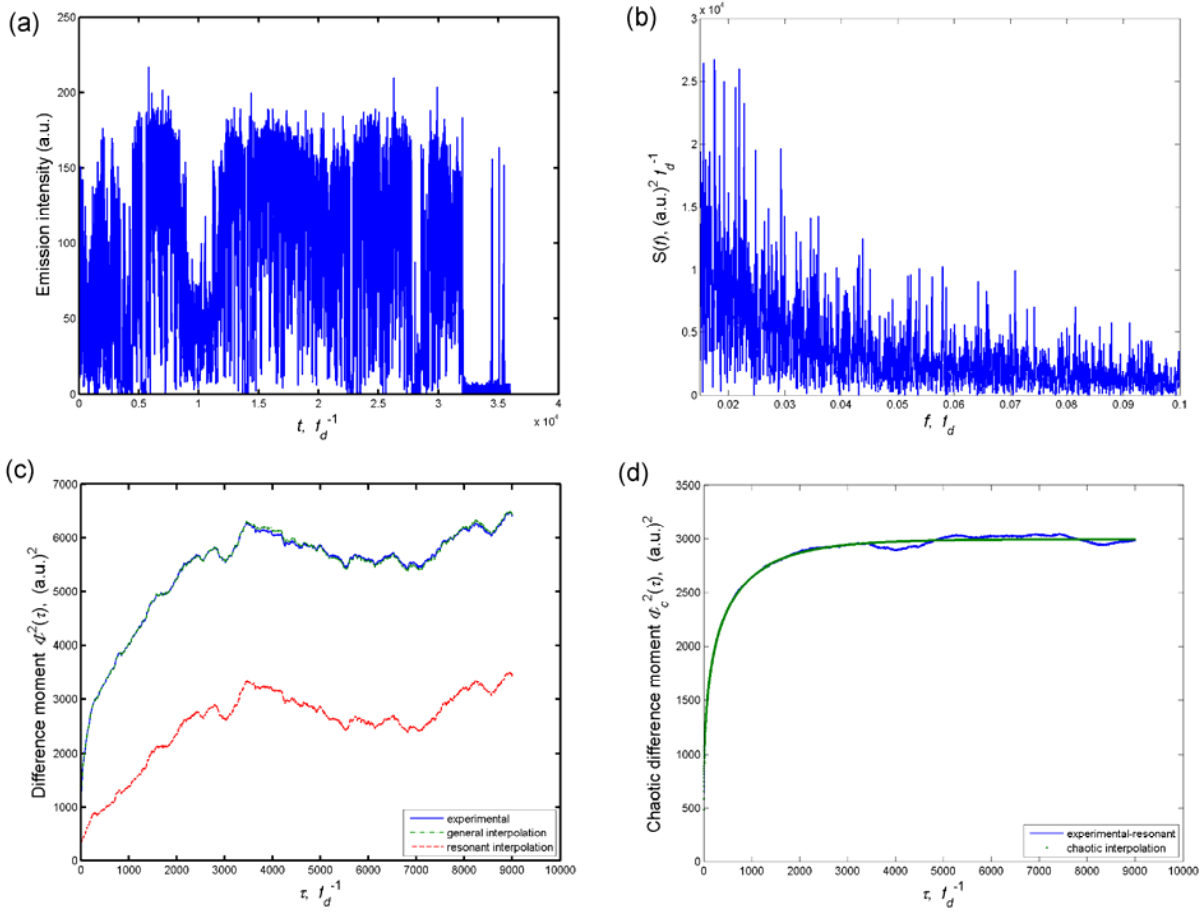

FIG. 10. Analysis the signal in Fig. 8 on the time interval from 162000 to 198000 $f_d^{-1}$ ($\sigma$ = 38.7 arb. un.; $H_1$ = 0.13; $T_1$ = 1585.7 $\Delta t \approx$ 15.86 s relative to the interval start, $D \approx$ 100.3 (arb. un.)$^2$/s, $S_c(0)$ = 9.35·10$^6$ (arb. un.)$^2 f_d^{-1}$; $n$ = 1.17; $T_0$ = 4028.3 $\Delta t \approx$ 40.3 s relative to the interval start; $q_{min}$ = 1 - see Appendix C).

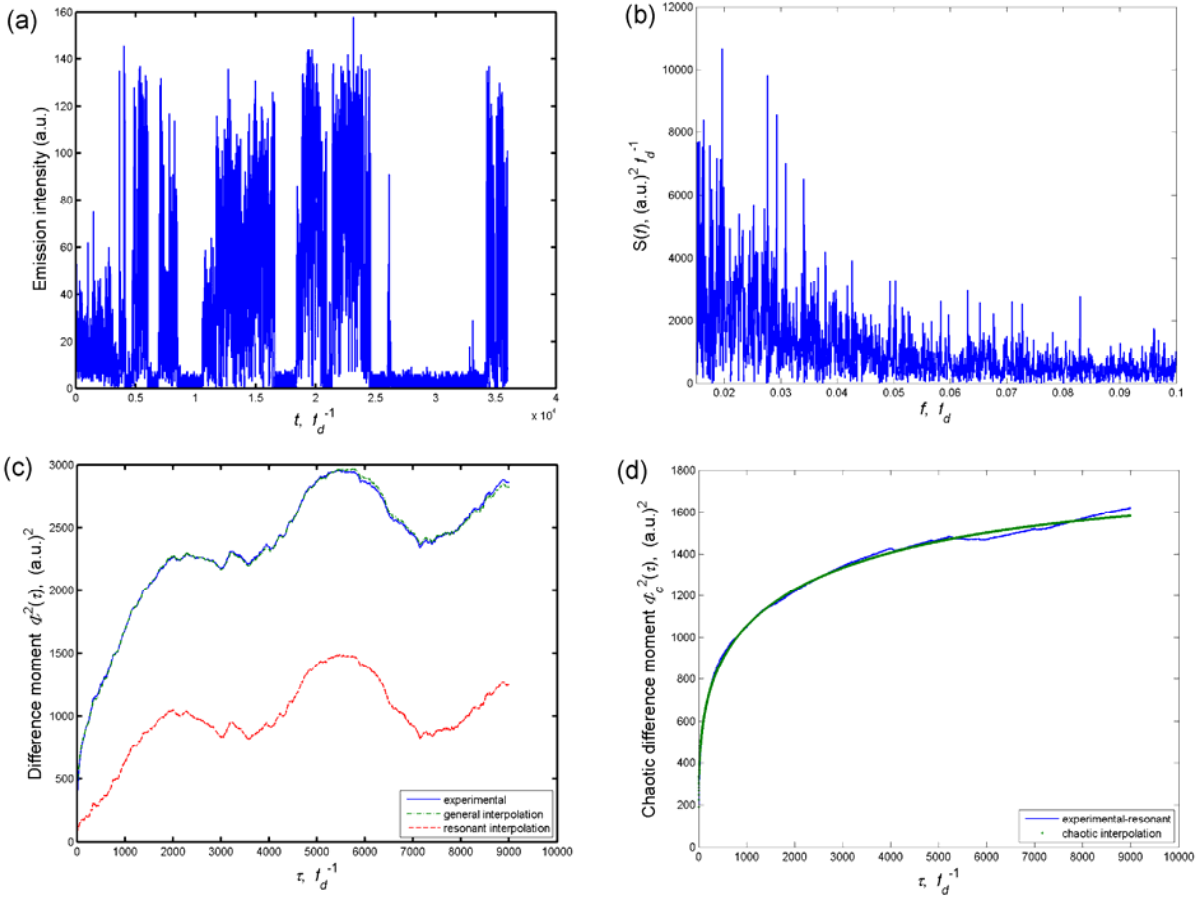

FIG. 11. Analysis the signal in Fig. 8 on the time interval from 324000 to 360000 $f_d^{-1}$ ($\sigma = 29.2$ arb. un.; $H_1 = 0.13$; $T_1 = 10076.6\ \Delta t \approx 100.1$ s relative to the interval start, $D \approx 8.96$ (arb. un.)$^2$/s, $S_c(0) = 3.88 \cdot 10^6$ (arb. un.)$^2 f^{-1}{}_d$; $n = 1.23$; $T_0 = 3546.5\ \Delta t \approx 35.5$ s relative to the interval start; $q_{min} = 1$ - see Appendix C).

### C. Anomalous diffusion in the dynamics of X-ray emission from accreting objects

This section shows that anomalous diffusion, as a process of stochastic state changes of a complex system, can manifest itself in the dynamics of stellar objects. Here, we will consider the example of the dynamics of X-ray emission from two accreting stellar systems: GRS 1915+105 and Cygnus X-1.

Binary system GRS 1915+105, which is located in the Aquila constellation approximately 40,000 light years away from the Sun, is a star-donor with mass $M_d = 1.2 \pm 0.2\ M_{sun}$, where $M_{sun}$ is the mass of the Sun, that rotates around a spinning compact heavy object, black hole with mass $M_{bh} = 14 \pm 4\ M_{sun}$.

The orbital period of this system is 33.5 ± 1.5 days. The interest of researchers to this system [38-43] was brought about by the high luminosity of its accretion disk, which is close to the Eddington limit, when the radiation pressure on the accreting matter is comparable to the gravitational attraction to the central object. This leads to unstable modes of matter transfer, which produce powerful X-ray flashes and gas streams (jets). For this reason, GRS 1915+105 is considered as a "microquasar", a stellar analog of active galactic nuclei [38].

The other microquasar, binary system Cygnus X-1, is a powerful source of X-ray emission, which is located 6,000 light years away from the Sun. The optical component of this system is a blue supergiant variable star with surface temperature around 31,000 K and the mass of 33 ± 9 $M_{sun}$. The lower limit of the mass of the accreting object, a black hole into which the matter flows from the atmosphere of the supergiant resulting in a flat gas disk, is estimated as 16 ± 5 $M_{sun}$.

X-ray emission (impulses with various powers and durations – up to milliseconds) is generated in the inner layers of the gas disk, the temperatures of which are estimated to be of the orders of $10^7$-$10^8$ K [40].

The time series $I(t)$ for the total flow of X-ray emission from these sources (the primary data are available on the Internet [44]) in the period from January 1, 1996 to December 31, 2005 are shown in Figs. 12(a) and 13(a). The average intervals between adjacent measurements were $\Delta t_{grs}$ = 106 minutes and $\Delta t_{cygx}$ = 88 minutes, with the corresponding sampling frequencies $f_d$ of $1.57\times 10^{-4}$ $Hz$ and $1.89\times 10^{-4}$ $Hz$; the total numbers of measurements were 49,355 and 59,748 for GRS 1915+105 and Cygnus X-1, respectively. The errors in the measurements were attributed to the variation of the intervals between adjacent measurements: approximately in 10% of the cases the intervals deviated from the average values $\Delta t_{grs}$ and $\Delta t_{cygx}$ by two times. The dynamics of the X-ray emission sources was previously studied in Refs. [39-41]. Based on the analysis of probability density functions and nonlinear dynamics of complex systems, it was suggested that the physical mechanisms of matter transfer are different between the systems and concluded that the self-similarity is limited in the dynamics of transfer [39, 40]. It was shown that the signals produced by Cygnus X-1 during the transfer of matter in the accretion disk could be characterized using a model of anomalous diffusion with the Hurst constant $H_1 \approx 0.3$ and time self-similarity on interval $T_s \approx 3$ years. Some of the features of transfer processes in the accretion disk of GRS 1915+105 were similar to the corresponding processes in Cygnus X-1: $H_1 \approx 0.35$, which also pointed to "subdiffusion". However, the time self-similarity was seen only on the intervals of $T_s \approx$ 12-17 days.

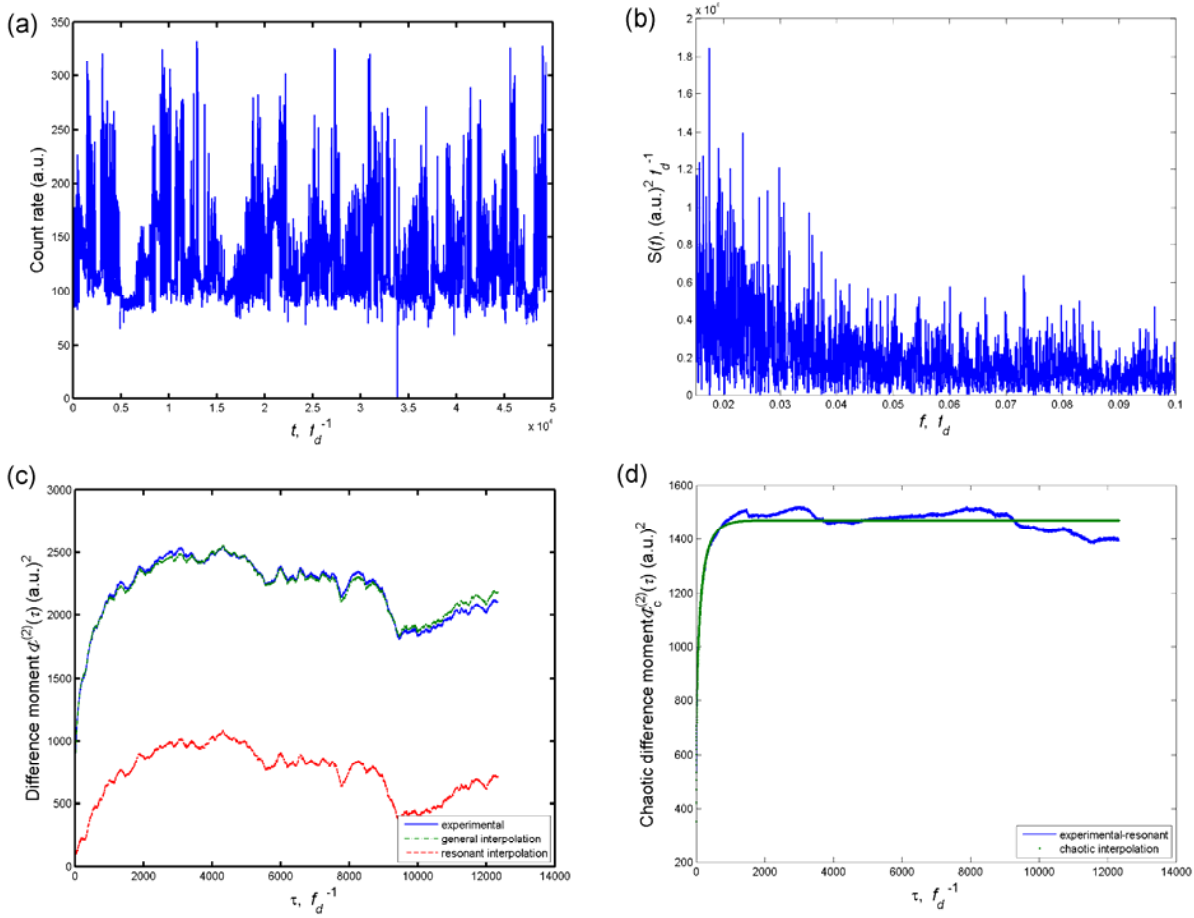

FIG. 12. Dynamics of X-ray emission from binary system GRS 1915+105 in the period from January 1, 1996 to December 31, 2005 [44] ($\sigma$ = 27.1 arb. un.; $H_1$ = 0.13; $T_1$ = 396 $\Delta t_{grs}$ ≈ 28.9 days, $D$ ≈ 27.0 (arb. un.)$^2$/day, $S_c(0)$ = 5.57·10$^5$ arb. un.; $n$ = 1.13; $T_0$ = 568.4 $\Delta t_{grs}$ ≈ 41.4 days): (a) source signal; (b) power spectrum $S(f)$ given by Eq. (4) in the low-frequency range (100 frequencies); (c) "experimental" – $\Phi^{(2)}(\tau)$ given by Eq. (5), "general interpolation" – $\Phi^{(2)}(\tau)$ given by Eq. (38), "resonant interpolation" – $\Phi_r^{(2)}(\tau)$ given by Eq. (37); (d) – "experimental – resonant" – $\Phi^{(2)}(\tau)$ given by Eq. (38) minus $\Phi_r^{(2)}(\tau)$ given by Eq. (37), "chaotic interpolation" – $\Phi_c^{(2)}(\tau)$ given by Eq. (7).

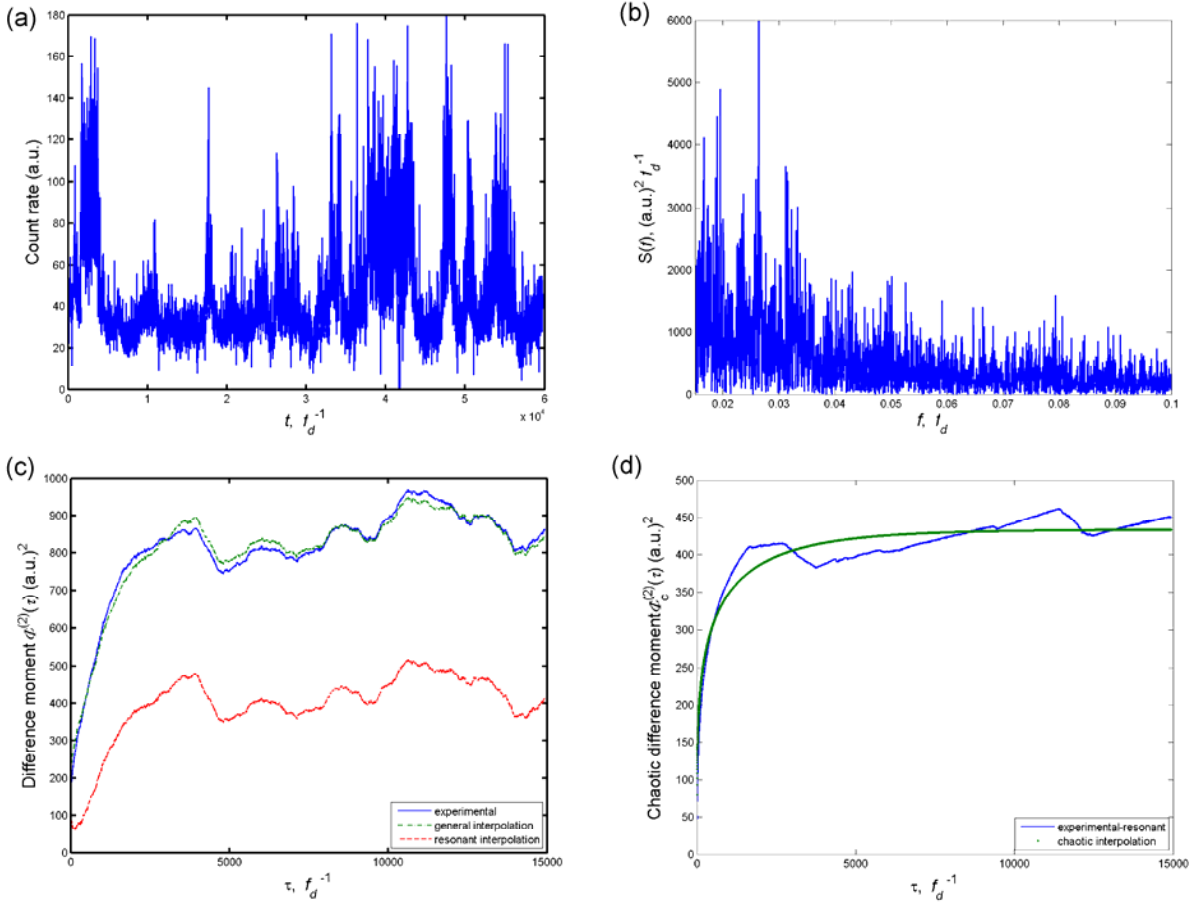

FIG. 13. Dynamics of X-ray emission from binary system Cygnus X-1 in the period from January 1, 1996 to December 31, 2005 [44] ($\sigma$ = 14.7 arb. un.; $H_1$ = 0.11; $T_1$ = 3442 $\Delta t_{cygx}$ ≈ 210 days, $D$ ≈ 1.09 (arb. un.)$^2$/day; $S_c(0)$ = 8.8·10$^5$ arb. un.; $n$ = 1.29; $T_0$ = 2123 $\Delta t_{cygx}$ ≈ 130 days): (a) source signal; (b) power spectrum $S(f)$ given by Eq. (4) in the low-frequency range (60 frequencies); (c) "experimental" – $\Phi^{(2)}(\tau)$ given by Eq. (5), "general interpolation" – $\Phi^{(2)}(\tau)$ given by Eq. (38), "resonant interpolation" – $\Phi_r^{(2)}(\tau)$ given by Eq. (37); (d) – "experimental – resonant" – $\Phi^{(2)}(\tau)$ given by Eq. (38) minus $\Phi_r^{(2)}(\tau)$ given by Eq. (37), "chaotic interpolation" – $\Phi_c^{(2)}(\tau)$ given by Eq. (7).

The results of FNS analysis of the signals shown in Figs. 12(a)-13(a) are presented in Figs. 12(b)-(d) and 13(b)-(d), with the FNS parameters being given in the figure captions. Though the actual values of $H_1$, 0.13 for GRS 1915+105 and 0.11 for Cygnus X-1, were different from the values given in Refs. [32, 33], 0.35 and 0.3, respectively, the FNS parameters pointed to the same mode of "subdiffusion". It should be noted that values of parameters $H_1$, $\sigma$, and $T_1$, calculated using the method of non-linear least squares, were chosen by providing the best agreement between the experimental and calculated curves $\Phi^{(2)}(\tau)$ in the entire interval of $\tau$ under study. At the same time, the Hurst constant $H_1$ in Refs. [39, 40] was calculated based on the agreement with experimental data in the limit of small values of $\tau$. More significant differences were seen for the values of $T_1$: 42 days for GRS 1915+105 and 210 days for Cygnus X-1. The differences between the values of $T_1$ and $T_s$ should be mostly attributed the fact that the methods of analysis used in Refs. [39, 40] do not include the separation of regular components from the signal before performing the chaotic parameterization. This factor was less important in calculating $H_1$, as only the small values of $\tau$ (high-frequency range) were used, where the effect of regular components is minimal. In calculating $T_s$, the low-frequency regular components play an important role and thus should be removed before performing the chaotic parameterization.

The above analysis shows the stochastic changes in the states of microquasars, i.e., astrophysical objects, can be described in terms of anomalous diffusion.

## VII. CONCLUSIONS

The above analysis demonstrates that anomalous diffusion manifests itself in the chaotic dynamics of magnetoencephalograms, blinking fluorescence of quantum dots, and X-ray emission from accreting objects, which are completely unrelated natural processes running at different scales: microscales for quantum dots and macroscales for stellar objects. This suggests that anomalous diffusion can be identified in many other natural signals.

As natural complex signals usually contain both regular and chaotic components, the calculation of the parameters of anomalous diffusion for natural stochastic signals should be performed after the regular components are removed from the source signals, just like it was done in this study.

The results of this study make it possible to model the chaotic dynamics of some natural processes. In the case of anomalous diffusion, the mathematical models should include fractional-derivative differential equations with the integrodifferential boundary conditions incorporating the effects of nonstationarity and the finite residence times of the diffusion system in boundary "adstates".

**ACKNOWLEDGEMENTS**

The authors are grateful to Professor Masaro Kuno (University of Notre Dame) for the quantum-dot experimental data used for the analysis in section VI.B. The authors would also like to thank S. A. Demin and O. Yu. Panischev for helpful discussions of the results concerning the analysis of magnetoencephalograms (section VI.A).

This study was supported in part by the Russian Foundation for Basic Research, project no. 08-02-00230 *a*.

**APPENDIX A**

Let us demonstrate that the information contents of $S_c(f)$ and $\Phi_c^{(2)}(\tau)$ coincide if there is no intermittence by considering the case of completely "irregular" dynamics of the Weierstrass – Mandelbrot (WM) function.

The real part of WM function is written as [45, 46]:

$$F_{WM}(t) = \sum_{n=-\infty}^{\infty} \frac{1-\cos b^n t}{b^{(2-D)n}} \qquad (b>1,\ 1<D<2).$$

Though continuous, this function cannot be differentiated at any point.

Autocorrelator $\psi_{WM}(\tau)$, transient difference moment of second order $\Phi_{WM}^{(2)}(\tau)$, and power spectrum $S_{WM}(f)$ for WM are expressed as [45]:

$$\psi_{WM}(\tau) = \langle F_{WM}(t)F_{WM}(t+\tau)\rangle = \frac{1}{2}\sum_{n=-\infty}^{\infty}\frac{\cos b^n\tau}{b^{2(2-D)n}},$$

$$\Phi_{WM}^{(2)}(\tau) = \langle [F_{WM}(t)-F_{WM}(t+\tau)]^2\rangle = \sum_{n=-\infty}^{\infty}\frac{1-\cos b^n\tau}{b^{2(2-D)n}},$$

$$S_{WM}(f) = \left|2\int_0^\infty F_{WM}(\tau)\cos(2\pi f\tau)d\tau\right|^2 = \frac{1}{4}\sum_{n=-\infty}^{\infty}\frac{\delta(2\pi f-b^n)}{b^{2(2-D)n}} \xrightarrow[2\pi(b-1)f<<1]{} \frac{1}{4\ln b\,(2\pi f)^{5-2D}}.$$

In this case, the constant ($\tau$-independent) term in the expression for $\psi_{WM}(\tau)$ was discarded. The corresponding term in the expression for power spectrum $S_{WM}(f)$, which characterizes the null frequency, was also discarded.

It is easy to show that functions $\Phi_{WM}^{(2)}(\tau)$ and $S_{WM}(f)$ can be expressed in terms of each other:

$$\psi_{WM}(\tau) = 2\int_0^\infty S_{WM}(f)\cos(2\pi f\tau) = \frac{1}{2}\sum_{n=-\infty}^{\infty}\frac{\cos b^n\tau}{b^{2(2-D)n}},$$

$$\Phi_{WM}^{(2)}(\tau) = 2[\psi_{WM}(0)-\psi_{WM}(\tau)] = \sum_{n=-\infty}^{\infty}\frac{1-\cos b^n\tau}{b^{2(2-D)n}},$$

$$S_{WM}(f) = 2\int_0^\infty \psi_{WM}(\tau)\cos(2\pi f\tau)d\tau = \frac{1}{4}\sum_{n=-\infty}^{\infty}\frac{\delta(2\pi f-b^n)}{b^{2(2-D)n}}.$$

Hence, the information contents of $\Phi_{WM}^{(2)}(\tau)$ and $S_{WM}(f)$ are the same despite the chaotic nature of the WM function.

**APPENDIX B**

Consider the diffusion problem given by (13), (15), (29) and (30):

$$\frac{\partial W}{\partial \tau} = D\frac{\partial^2 W}{\partial V^2} \tag{13}$$

$$W(V,0) = \delta(V). \tag{15}$$

$$D\frac{\partial W}{\partial V} = \chi W(-L,\tau) - \lambda\chi \exp(-\lambda\tau)\int_0^\tau W(-L,\xi)\exp(\lambda\xi)d\xi \quad \text{при } V = -L, \tag{29}$$

$$-D\frac{\partial W}{\partial V} = \chi W(L,\tau) - \lambda\chi \exp(-\lambda\tau)\int_0^\tau W(L,\xi)\exp(\lambda\xi)d\xi \quad \text{при } V = L. \tag{30}$$

We will use an iterative method for solving the problem. At the initial step, we will approximate $W(L,\xi)$ and $W(-L,\xi)$ in the boundary conditions with the expression

$$W(V,\tau) = \frac{1}{2L}\left[1 + 2\sum_{k=1}^\infty \exp\left(-\frac{\pi^2 k^2 D\tau}{L^2}\right)\cos\frac{\pi k V}{L}\right], \tag{16}$$

which was obtained for the case of the boundary conditions of symmetry (14).

At the boundaries, expression (16) takes the following form:

$$W(-L,\tau) = W(L,\tau) = \frac{1}{2L}\left[1 + 2\sum_{k=1}^\infty (-1)^k \exp\left(-\frac{\pi^2 k^2 D\tau}{L^2}\right)\right]. \tag{B.1}$$

Considering the facts that the first term of the sum approximates well the total sum for most of the time interval because $\exp(-k^2)$ dramatically decreases and that the value of $W$ is close to zero at the initial time interval, we can use the following approximation:

$$W(-L,\tau) = W(L,\tau) \approx \frac{1}{2L} \times U\left[1 - 2\exp\left(-\frac{\pi^2 D\tau}{L^2}\right)\right] \times \left\{1 - 2\exp\left(-\frac{\pi^2 D\tau}{L^2}\right)\right\}. \tag{B.2}$$

Here, $U\left[1 - 2\exp\left(-\frac{\pi^2 D\tau}{L^2}\right)\right]$ is the Heaviside function, which is equal to "0" if the argument is less than "0" and is equal to "1" if the argument is "0" or more than "0".

Comparison of the analytical expression (B.2) and numerical solution, which was obtained for the symmetry boundary conditions (14) using the *pdepe* function, a built-in MATLAB numerical procedure to solve initial-boundary problems for one-dimensional parabolic-elliptic partial differential equations, with properly selected coordinate step to match the initial condition based on the Dirac delta function, it can be seen that the approximation is in good agreement with the numerical solution (see Fig. B.1).

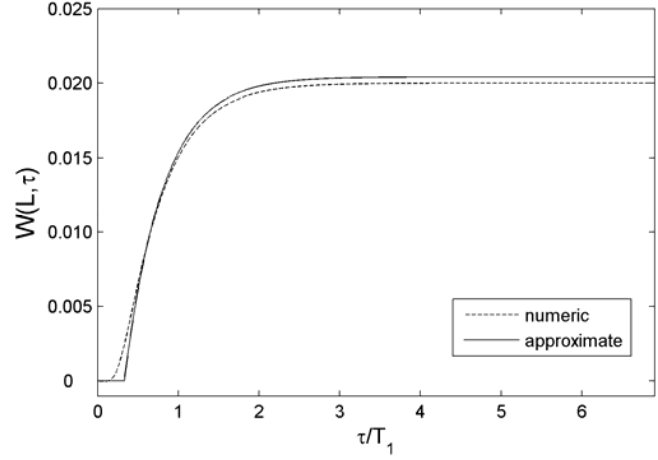

FIG. B.1. Comparison of Eq. (B.2) with the numerical solution for boundary conditions (14).

By evaluating the integrals in boundary conditions (29) and (30) with $W(L,\xi)$ and $W(-L,\tau)$ given by Eq. B.2, we obtain

$$D\frac{\partial W}{\partial V} = \chi W(-L,\tau) - \frac{\lambda \chi \exp(-\lambda \tau)}{2L} \times U\left[1 - 2\exp\left(-\frac{\pi^2 D\tau}{L^2}\right)\right] \times \Theta(\tau), \qquad (B.3)$$

$$-D\frac{\partial W}{\partial V} = \chi W(L,\tau) - \frac{\lambda \chi \exp(-\lambda \tau)}{2L} \times U\left[1 - 2\exp\left(-\frac{\pi^2 D\tau}{L^2}\right)\right] \times \Theta(\tau) \qquad , \qquad (B.4)$$

where

$$\Theta(\tau) = \left[\left\{\frac{\exp(\lambda \tau) - \exp\left(\frac{\ln(2)\lambda L^2}{\pi^2 D}\right)}{\lambda}\right\} - 2\frac{\exp\left(\left\{\lambda - \frac{\pi^2 D}{L^2}\right\}\tau\right) - \exp\left(\left\{\lambda - \frac{\pi^2 D}{L^2}\right\}\frac{\ln(2)L^2}{\pi^2 D}\right)}{\lambda - \frac{\pi^2 D}{L^2}}\right].$$

To solve problem (13), (15), (29) and (30), we first numerically solved problem (13), (15), (B.3), and (B.4) using the *pdepe* function in MatLab, and then substituted the solution for $W(L,\tau)$ and $W(-L,\tau)$ into the boundary conditions (29) and (30). The resulting problem was solved numerically. The procedure kept iterating until the required accuracy for $W(L,\tau)$ was achieved. The variations of $W$ and structural function at the boundaries for the initial, first, and last iterations are shown in Fig. B.2.

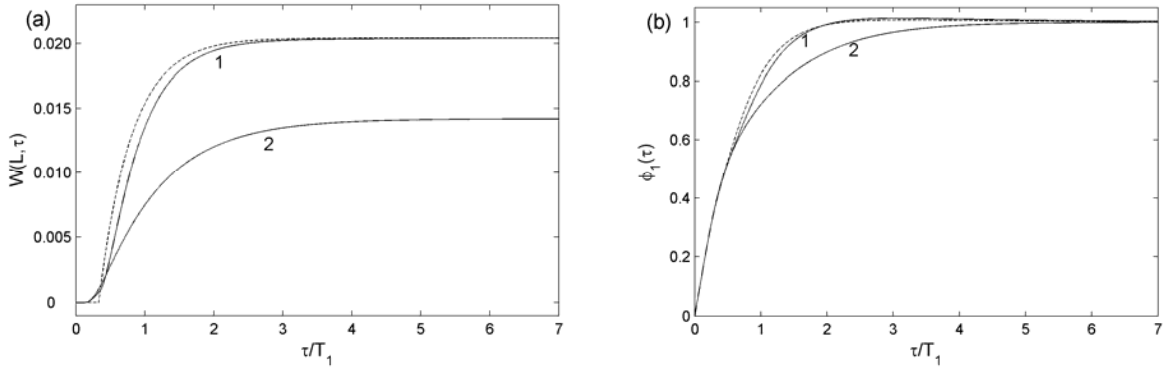

FIG. B.2. Functions $W(L,\tau)$ (a) and $\phi_1(\tau)$ (b) for the initial (dashed), first (curve 1), and last (curve 2) iterations.

**APPENDIX C**

PARAMETERIZATION ALGORITHM (customized for MATLAB)

Notation

$q_{min}$ is the number of frequency points to ignore when calculating $S_c(0)$ [step 3] and estimating resonant (regular) power spectrum [step 5];

$q_{max}$ is the highest frequency point used in calculating $S_c(0)$ [step 3];

Default values:

$$T_M = \frac{T}{4}, \; q_{min} = 0, \; q_{max} = q_{min} + 1.$$

STEP 1. Subtract the mean from the original signal.

STEP 2. Calculate the discrete cosine transform of the autocorrelator.

Calculate the autocorrelator:

$$\psi(m) = \frac{1}{N-m} \sum_{k=1}^{N-m} V(k)V(k+m) \quad \text{for } 0 \leq m < M,$$

where

$T_M \leq \frac{T}{2}$, $M = \left\lfloor \frac{T_M}{T} N \right\rfloor$ is the number of points on the frequency axis, and $N$ is the number of points in the averaging interval.

Fill in $\psi(m)$ for $m=M+1..2M-1$ using $\psi(m) = \psi(2M-m)$, as the autocorrelator is symmetric. In other words, we use only the first M+1 (0..M) values, which corresponds to our "interval of interest" $T_M$. Other values are obtained from the symmetry condition.

Calculate the Fast Fourier transform of the autocorrelator:

$$S_{exp}(q) = \sum_{m=0}^{2M-1} \psi(m) \exp\left(-\frac{\pi i\, qm}{M}\right),$$

where $f$ is the frequency, $q = 2f T_M$, and $S_{exp}(f) = \Delta t \times S_{exp}(q)$.

For real signals, $S_{exp}(q) = S_{exp}(2M-q)$. So, keep only the first M+1 frequency points of $S_{exp}$.

For $q = 1..M-1$, multiply the values of $S_{exp}(q)$ by 2 (go to cosines from complex Fourier).

Result: $S_{exp}(q)$ for $q = 0..M$.

STEP 3. Calculate $S_c(0)$ using $S_{exp}$.

Choose a specific frequency range $[q_{min}, q_{max}]$ in the range of low frequencies. Find the minimum value of $S_{exp}$ in the range and set $S_c(0)$ to this value. Formally,

$$S_c(0) = \min_{q \in [q_{min},\, q_{max}]} S_{exp}.$$

STEP 4. Interpolate $S_{exp}(q)$ [q>0] with $S_c(q) \approx \dfrac{S_c(0)}{1 + (2\pi \frac{q}{T_M} T_0)^n}$ to find parameters $n$ and $T_0$.

Use the nonlinear method of least squares.

STEP 5. Calculate $S_r(q) = S_{exp}(q) - S_c(q)$. For $q = 0..q_{min} - 1$, set $S_r(q) = 0$.

STEP 6. Calculate the autocorrelator for the resonant (regular) component.

For $q = 2..M - 1$, set $S_r = S_r / 2$.

Use the symmetry complement to fill in $S_r$ for m=M+1..2M-1: $S_r(q) = S_r(2M - q)$.

Calculate the resonant autocorrelator

$$\psi_r(m) = \frac{1}{2M} \sum_{q=0}^{2M-1} S_r(q) \exp\left(\frac{\pi i q m}{M}\right).$$

Keep only the first $M + 1$ points.

STEP 7. Calculate the difference moment for the resonant component:

$$\Phi_r^{(2)}(m) = 2[\psi_r(0) - \psi_r(m)] \text{ for } m = 0..M.$$

STEP 8. Calculate the difference moment for experimental series:

$$\Phi^{(2)}(m) = \frac{1}{N-m} \sum_{k=1}^{N-m} [V(k) - V(k+m)]^2 \text{ for } m = 0..M.$$

STEP 9. Calculate the difference moment for the chaotic component:

$$\Phi_c^{(2)}(m) = \Phi^{(2)}(m) - \Phi_r^{(2)}(m) \text{ for } m = 0..M.$$

STEP 10. Interpolate the chaotic difference moment using the function

$$\Phi_c^{(2)}(m) = 2\sigma^2 \cdot \left[1 - \Gamma^{-1}(H_1) \cdot \Gamma(H_1, m \times \Delta t / T_1)\right]^2,$$

$$\Gamma(s,x) = \int_x^\infty \exp(-t) \cdot t^{s-1} dt, \quad \Gamma(s) = \Gamma(s,0).$$

Find parameters $\sigma$, $H_1$, $T_1$.

STEP 11. Calculate the jump component of the power spectrum

$$S_{cJ}(q) \approx \frac{S_{cJ}(0)}{1 + (2\pi \frac{q}{T_M} T_1)^{2H_1+1}},$$

where

$$S_{cJ}(0) = 4\sigma^2 T_1 H_1 \cdot \left\{ 1 - \frac{1}{2H_1 \Gamma^2(H_1)} \int_0^\infty \Gamma^2(H_1, \xi) d\xi \right\}.$$